\documentclass{emulateapj}

\newcommand{\Kepler}{{\it Kepler}}

\newcommand{\vespa}{{\texttt{vespa}}}

\newcommand{\rearth}{R$_\oplus$}

\newcommand{\meh}{[M/H]}

\newcommand{\ldoneg}{0.55}
\newcommand{\uldoneg}{0.14}
\newcommand{\ldtwog}{0.447}
\newcommand{\uldtwog}{0.045}
\newcommand{\rprstg}{0.0153}
\newcommand{\urprstg}{0.0019}
\newcommand{\arstg}{45}
\newcommand{\uarstg}{^{+9.9}_{-15}}
\newcommand{\inclg}{89.35}
\newcommand{\uinclg}{^{+0.47}_{-0.98}}
\newcommand{\impg}{0.52}
\newcommand{\uimpg}{0.28}
\newcommand{\rplg}{1.13}
\newcommand{\urplg}{0.14}
\newcommand{\perplg}{14.64558}
\newcommand{\uperplg}{0.00012}
\newcommand{\ttransitg}{2455658.6073}
\newcommand{\uttransitg}{0.0037}
\newcommand{\tdurg}{2.12}
\newcommand{\utdurg}{0.28}
\newcommand{\teqg}{418}
\newcommand{\uteqg}{36}

\newcommand{\ldonei}{0.42}
\newcommand{\uldonei}{0.13}
\newcommand{\ldtwoi}{0.278}
\newcommand{\uldtwoi}{0.040}
\newcommand{\rprsti}{0.00935}
\newcommand{\urprsti}{0.00089}
\newcommand{\arsti}{33.8}
\newcommand{\uarsti}{^{+6.6}_{-12}}
\newcommand{\incli}{89.2}
\newcommand{\uincli}{^{+0.59}_{-1.3}}
\newcommand{\impi}{0.50}
\newcommand{\uimpi}{0.28}
\newcommand{\rpli}{1.32}
\newcommand{\urpli}{0.21}
\newcommand{\perpli}{14.44912}
\newcommand{\uperpli}{0.00020}
\newcommand{\ttransiti}{2455644.3488}
\newcommand{\uttransiti}{0.0048}
\newcommand{\tduri}{2.80}
\newcommand{\utduri}{0.31}
\newcommand{\teqi}{709}
\newcommand{\uteqi}{75}

\usepackage{color}
\newcommand{\ron}{\color{black}}

\usepackage{hyperref}
\usepackage{breakurl}
\usepackage{amsmath}
\usepackage{textcomp}
\usepackage{verbatim}
\usepackage{graphicx}
\usepackage{longtable}
\slugcomment{}

\shorttitle{Identifying Exoplanets with Deep Learning}
\shortauthors{Shallue \& Vanderburg}

\begin{document}


%\title{Identifying Exoplanets with Deep Learning:\\ \ron The Discoveries of Kepler-80~\MakeLowercase{g} and Kepler-90~\MakeLowercase{i}, the Eighth Planet in its System}

\title{Identifying Exoplanets with Deep Learning: \ron A Five Planet Resonant Chain \\ around Kepler-80 and an Eighth Planet around Kepler-90}


\author{Christopher J. Shallue\altaffilmark{$\dagger$ 1} \& Andrew Vanderburg\altaffilmark{$\star$, 2,3}}

%\affil{Harvard--Smithsonian Center for Astrophysics, 60 Garden St., Cambridge, MA 02138}

\altaffiltext{$\dagger$}{\url{shallue@google.com}}
\altaffiltext{1}{Google Brain, 1600 Amphitheatre Parkway, Mountain View, CA 94043}
\altaffiltext{2}{Department of Astronomy, The University of Texas at Austin, 2515 Speedway, Stop C1400, Austin, TX 78712}
\altaffiltext{3}{Harvard--Smithsonian Center for Astrophysics, 60 Garden St., Cambridge, MA 02138}
\altaffiltext{$\star$}{National Science Foundation Graduate Research Fellow and NASA Sagan Fellow}


\begin{abstract}

NASA's {\it Kepler Space Telescope} was designed to determine the frequency of Earth-sized planets orbiting Sun-like stars, but these planets are on the very edge of the mission's detection sensitivity. Accurately determining the occurrence rate of these planets will require automatically and accurately assessing the likelihood that individual candidates are indeed planets, even at low signal-to-noise ratios. We present a method for classifying potential planet signals using deep learning, a class of machine learning algorithms that have recently become state-of-the-art in a wide variety of tasks. We train a deep convolutional neural network to predict whether a given signal is a transiting exoplanet or a false positive caused by astrophysical or instrumental phenomena. Our model is highly effective at ranking individual candidates by the likelihood that they are indeed planets: 98.8\% of the time it ranks plausible planet signals higher than false positive signals in our test set. We apply our model to a new set of candidate signals that we identified in a search of known \Kepler\ multi-planet systems. We statistically validate two new planets that are identified with high confidence by our model. One of these planets is part of a five-planet resonant chain around \Kepler-80, with an orbital period closely matching the prediction by three-body Laplace relations. The other planet orbits \Kepler-90, a star which was previously known to host seven transiting planets. Our discovery of an eighth planet brings \Kepler-90 into a tie with our Sun as the star known to host the most planets.

\end{abstract}


\section{Introduction}

{\ron Over the course of its prime mission from May 2009 to May 2013, NASA's {\it Kepler Space Telescope}} observed about 200,000 stars photometrically with unprecedented precision \citep{koch, jenkinslongcadence, christiansencdpp}, discovered thousands of transiting exoplanets
\citep[and thousands more planet candidates,][]{koi1,koi2, koi3, koi4, koi5, koi6,koi7,koi8,rowe, morton16}, and significantly improved our understanding of the population of small planets in the inner parts of planetary systems \citep{marcymasses, howard, youdin, dressingk93, wolfgang}.

\Kepler\ was designed as a statistical mission, with the goal of determining the occurrence rate of Earth-sized planets in temperate orbits around Sun-like stars -- that is, planets that might (in ideal circumstances) support life as we know it on Earth. But early in the mission, most of the major results coming from \Kepler\ were the discoveries of individual particularly interesting planets or planetary systems. \Kepler\ discovered the smallest transiting planets \citep{muirhead, barclay}, the closest transiting planets \citep{k10, jackson, ofir, sanchisojeda1, sanchisojeda2}, the most-distant transiting planets \citep{kippingjupiter, jiwangplanethunters, foremanmackeylongperiod}, and some of the most interesting and unusual systems in terms of orbital architectures \citep{lissauerk11, doylek16, cabrerak90}. As such, the first lists of planet candidates \citep{koi1, koi2, koi3} produced by the \Kepler\ team were assembled in a heterogeneous way in an effort to include as many planet candidates as possible. Most of the first lists of planet candidates were produced from lists of ``threshold crossing events'' (TCEs; detected periodic signals that may be consistent with transiting planets) from the \Kepler\ pipeline \citep{jenkins-overview, jenkins-tps}, which were manually culled by humans to remove false positives caused by astrophysical variability and instrumental artifacts. A few candidates came from other sources as well; \citet{koi3} listed four planet candidates identified by eye by volunteers for the Planet Hunters project \citep{fischer}.
% TODO: there are about 5 papers people cite for the Kepler pipeline, most cite 1-3. I cited 2 the most relevant? But not the most recent.

As the \Kepler\ mission matured and the list of planet candidates grew, the focus of the community shifted towards population-level studies \citep{fressin, petigura1, petigura2, foremanmackeyocc, dressingcharbonneau1, dressingcharbonneau2, burke, mulders1,mulders2}. This shift in focus also manifested in changing attitudes towards the way planet candidate lists were produced and vetted. While heterogeneous catalogs that rely on human judgment are an efficient way to spur follow-up observations of \Kepler's numerous exciting planet candidates, they are not well suited for population studies, which rely on uniformity. Uniform and automatically produced catalogs make it possible to characterize both {\ron precision} (the fraction of detections that are true planets) and {\ron recall} (the fraction of true planets that are detected), which are essential quantities for estimating population-level occurrence rates. Mid-way through the mission, efforts were taken to produce planet candidate catalogs in an automated way, relying less and less on human judgment in what became a planet candidate. Projects such as the Robovetter \citep{koi7}, a decision tree designed to mimic the manual process for rejecting false positive TCEs, progressed to the point that they could be used in lieu of human vetting to produce fully automated catalogs \citep[][]{koi7, autovettercat, koi8}.

Meanwhile, others began to explore the use of machine learning for automatically vetting \Kepler\ TCEs. The Autovetter project \citep{autovetter} used a random forest model to classify TCEs based on features derived from \Kepler\ pipeline statistics. A similar approach was presented by \citet{mislis}. \cite{lpp} and \cite{armstrong} used unsupervised machine learning to cluster \Kepler\ light curves with similar shapes, and defined classification metrics using the distances between new light curves and TCEs with known labels. The \textit{LPP} metric defined by \cite{lpp} is used by the Robovetter to filter out light curves with ``not transit like'' shapes. In addition to using machine learning on transiting planets in \Kepler, others have used machine learning to identify different types of candidate signals in \Kepler\ and other datasets. \citet{millholland} used supervised learning to identify candidate {\em non}-transiting planets in \Kepler\ data, and \citet{dittmann} used a neural network to identify the most likely real transits among many candidate events in the MEarth dataset. 

In this paper we present a deep neural network for automatically vetting \Kepler\ TCEs. Our model uses light curves as inputs and is trained on a set of human-classified \Kepler\ TCEs. Neural networks have previously been applied to a variety of astronomical problems. \cite{pearson} used a neural network to detect planet transits in simulated light curves. Our best model has similarities with those of \cite{cabrera} for transient detection and \cite{schaefer} for strong gravitational lens detection. Like those papers, we feed views of our inputs through separate convolutional columns, a technique that has previously been successful for image classification \citep{ciresan}. Our model is able to distinguish with good accuracy the subtle differences between genuine transiting exoplanets and false positives like eclipsing binaries, instrumental artifacts, and stellar variability.

This paper is organized as follows. First, in Section \ref{neuralnet}, we introduce the concept of neural networks and describe several different neural network architectures that may be useful for problems like vetting transiting planet candidates. We describe the inputs to our neural network models in Section \ref{trainingset}, and describe how we designed and trained our models in Section \ref{automaticvetting}. In Section \ref{analysis}, we analyze the performance of our models and visualize some of the features our models learned about transiting planets and false positive signals. In Section \ref{newcandidates}, we use our best model to rank TCEs from a new search of stars observed by \Kepler\ that are already known to host multiple transiting planet candidates. We discover several new planet candidates, and apply further scrutiny to statistically validate the best few candidates. In Section \ref{discussion}, we discuss two newly discovered planets -- \Kepler-80~g and \Kepler-90~i -- and we also discuss the prospects for using our technique for occurrence rate calculations in the future.

\section{Deep Neural Networks}\label{neuralnet}

\subsection{Introduction}

The goal of machine learning is to combine input features into useful outputs. For example, we might wish to build an image classification model that learns to assign labels (e.g. ``dog'', ``car'', ``tree'') to input images. To build this model, we could manually define the input features that our model needs to make classification decisions (e.g. ``number of legs'' and ``number of wheels'' would be useful features for distinguishing between dogs and cars). However, defining those features in terms of image pixels is extremely difficult. Instead, we could build a model that learns features automatically from data: this is called \textit{representation learning}.

\textit{Deep learning}, which is a type of representation learning, uses computational \textit{layers} to build increasingly complex features that are useful -- in particular -- for classification problems \citep{nnnature}. For example, a deep image classification model might first detect simple edge features, which can then be used to detect curves and corners, and so on, until the model's final feature layer can discriminate between complex objects. Deep \textit{neural networks}, which are a type of deep learning model, have recently become the state-of-the art in a variety of tasks \citep[e.g. image classification,][]{alexnet}, and often outperform models built with hand-designed features.

\subsection{Fully Connected Neural Networks}
\begin{figure}[t!]
\centering
\includegraphics[width=0.8\columnwidth]{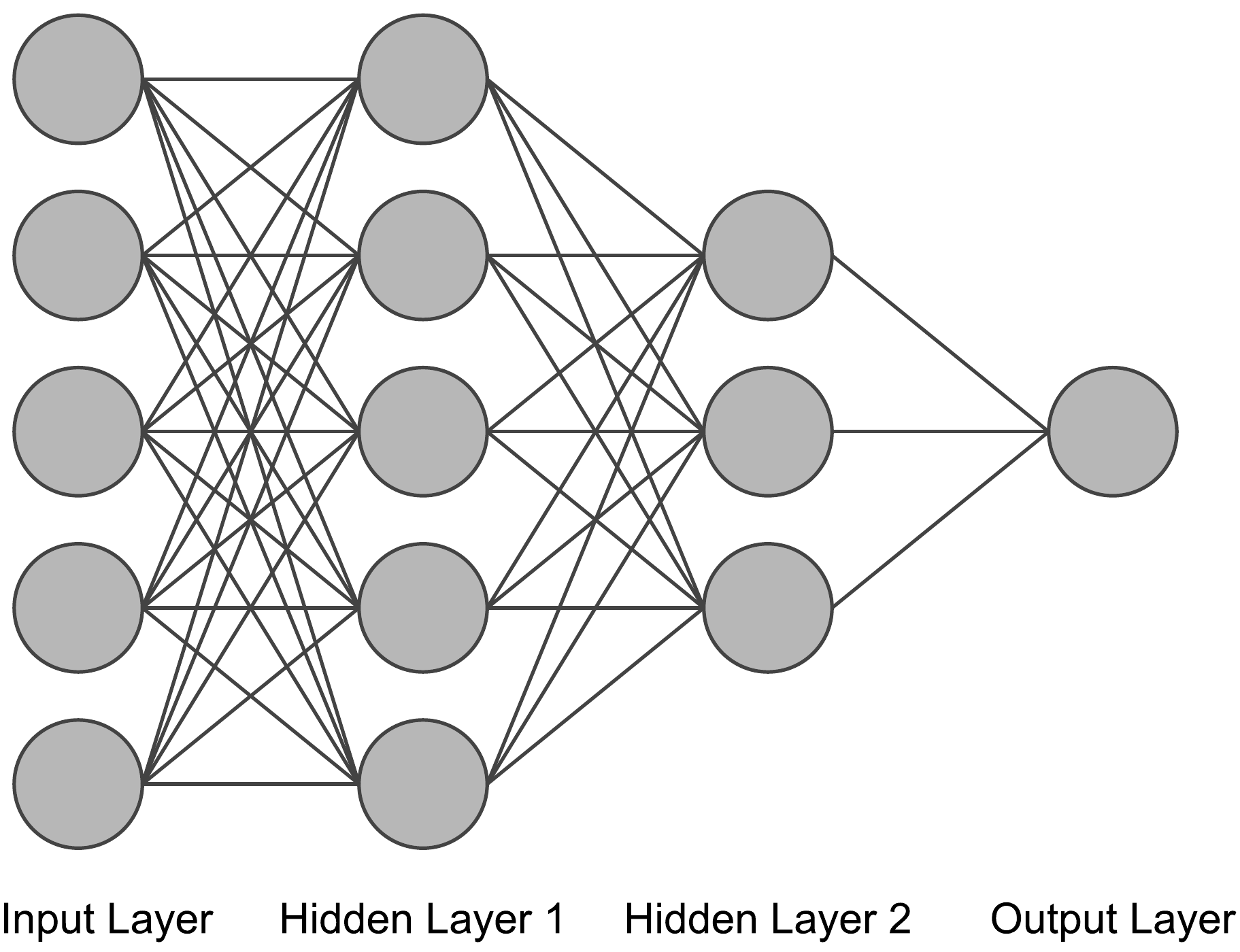}
\caption{A fully connected neural network. Each connection represents a multiplicative weight parameter learned by the model. Inputs are fed into the first layer of the network, the output layer generates predictions, and the hidden layers represent a hierarchy of learned features.} \label{fcnn}
\end{figure}

Figure \ref{fcnn} shows a \textit{fully connected neural network} (also known as a \textit{multilayer perceptron} or \textit{feed-forward neural network}). Its layers, which are comprised of scalar-valued units called \textit{neurons}, are arranged hierarchically: the outputs from one layer are the inputs to the next. The output value of a neuron for a specific set of inputs is called its \textit{activation}. The activations of the first layer (the \textit{input layer}) are the inputs to the network, and the activations of the final layer (the \textit{output layer}) are the network's predictions given those inputs. All other layers are \textit{hidden layers} because their activations are not directly observed.

Hidden layer activations are defined by the equation:
{\ron
\begin{equation}
\mathbf{a}_i = \phi (\mathbf{W}_i \mathbf{a}_{i-1} + \mathbf{b}_i )
\end{equation}

\noindent where $\mathbf{a}_i$ is the vector of length $n_i$ of activations in layer $i$, $\mathbf{W}_i$ is an $n_i \times n_{i-1}$ matrix of learned \textit{weight} parameters, $\mathbf{b}_i$ is a vector of length $n_{i}$ of learned \textit{bias} parameters, and $\phi$ is an elementwise \textit{activation function}.
}
The activation function $\phi$ is always nonlinear (a stack of linear layers would be equivalent to a single linear layer), so it is also called a \textit{nonlinearity}. The optimal choice of $\phi$ is application dependent and is often determined experimentally. Popular choices include the sigmoid function $\phi(x)=1/(1+e^{-x})$, hyperbolic tangent $\phi(x) = \tanh(x)$, and the linear rectification function $\phi(x) = \max\{0, x\}$. The former two choices are \textit{saturating} activation functions because their derivatives approach zero as $x \rightarrow \pm \infty$, which can cause gradient-based optimization algorithms to converge slowly. The derivative of the linear rectification function does not diminish as $x \rightarrow + \infty$, which helps avoid this ``vanishing gradient problem'' and often results in faster training and better performance \citep{relu}. Neurons with this activation function are called \textit{rectified linear units} (ReLU).

For a classification problem, the output layer is normalized so that each neuron's value lies in $(0, 1)$ and represents the probability of a specific output class. A binary classification problem (i.e. two output classes, $\{0, 1\}$) typically has a single output neuron representing the probability of the positive class. In this case, it is standard to use the sigmoid activation function for the output neuron because its range is $(0,1)$ and it induces a particularly convenient formula for gradient-based optimization\footnote{The sigmoid is the inverse of the ``canonical link function'' of the Bernoulli distribution. Assuming the model represents a conditional Bernoulli distribution of the target variable given the input, the inverse of the canonical link function induces a simple formula for the derivative of the log likelihood of the training data with respect to the model's parameters \citep[][\S 4.3.6]{bishop}.}.

% The canonical link function of the Bernoulli distribution is the inverse sigmoid function (a link function is the inverse of an activation function)

% For binary classification, it is standard to denote the classes $\{0, 1\}$ and predict the probability of the positive class with a single output neuron that uses the sigmoid activation function

% For binary classification, it is standard to denote the classes $\{0, 1\}$ and use a single output neuron representing the probability of the positive class. In this case, the output neuron typically uses the sigmoid activation function because it induces a particularly convenient update rule for gradient-based optimization

% The sigmoid function is the inverse of the ``canonical link function'' of the Bernoulli distribution.

% A binary classification problem (i.e. two output classes, $\{0,1\}$) is naturally modelled by a neural network as a Bernoulli distribution with a single output neuron representing the probability of the positive class

\subsection{Convolutional Neural Networks}

\begin{figure}[t!]
\centering
\includegraphics[width=0.8\columnwidth]{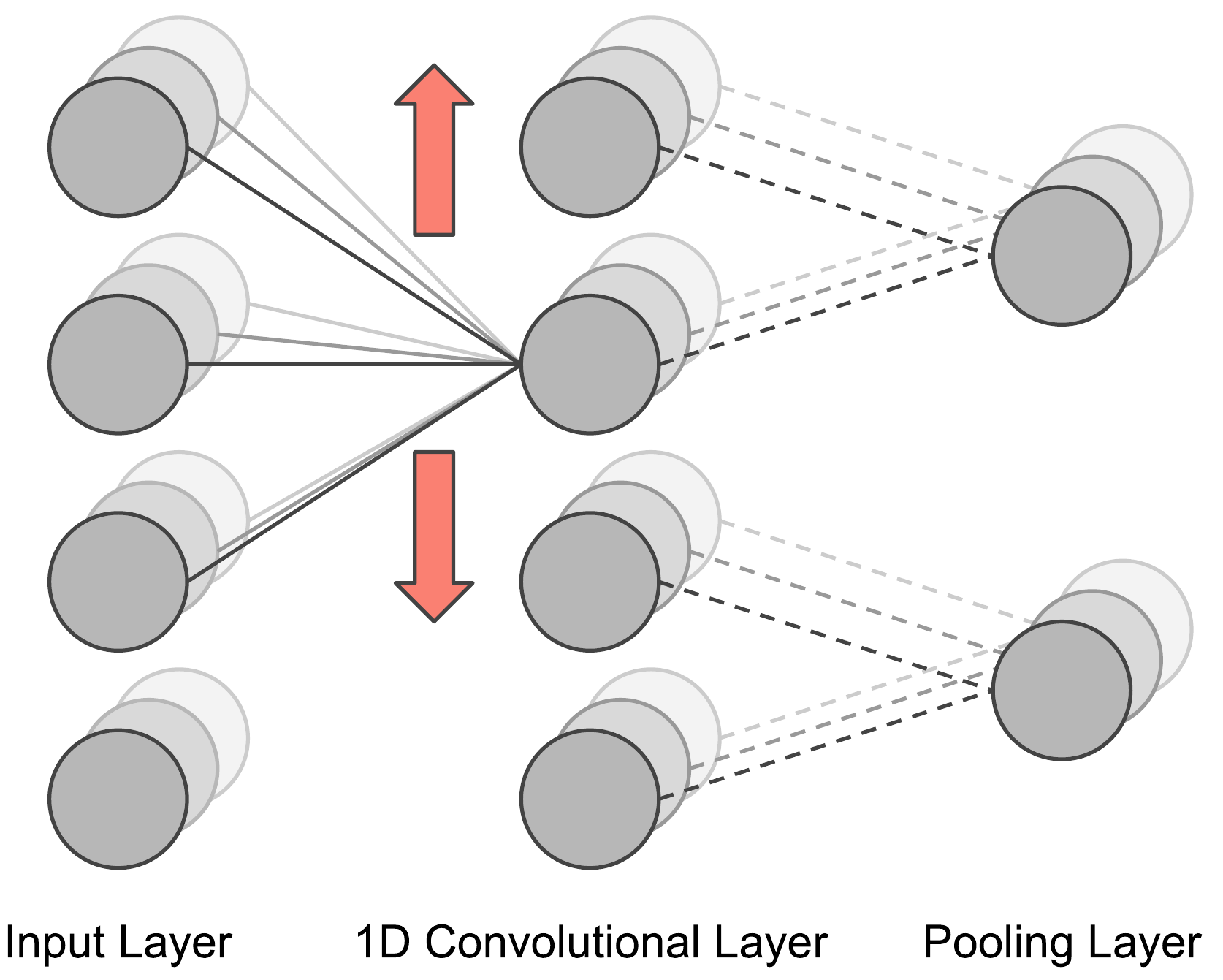}
\caption{Three layers of a 1-dimensional convolutional neural network. The convolutional layer takes the discrete cross-correlation operation of the vectors in its input layer with kernel vectors that are learned by the model. The pooling layer aggregates values in small neighborhoods along its input, typically by taking the mean or maximum value within each neighborhood. The pooling layer aggregates regions spaced $s$ neurons apart, where $s$ is the \textit{stride length}; here, the stride length is 2.} \label{cnn}
\end{figure}

Fully connected neural networks, which densely connect every neuron in layer $n$ to every neuron in layer $n+1$, ignore any spatial structure present in the input. For example, every pixel in a 2D input image would be treated independently without using the fact that some pixels are located near each other. Features that are composed of neighboring pixels, like edges and shapes, need to be learned independently by the model for each location in the input. By contrast, \textit{convolutional neural networks} (CNNs) exploit spatial structure by learning local features that are detected across the entire input; each feature only needs to be learned once. This significantly reduces the number of parameters that the model needs to learn, and reduces the memory usage and number of computational operations required to compute the output.

A CNN typically consists of \textit{convolutional layers} and \textit{pooling layers} (Figure \ref{cnn}). The input to a (1-dimensional) convolutional layer is a stack of {\ron $K$ vectors $\mathbf{a}_{i-1}^{(k)}$ $(k=1,2,...,K)$ of length $n_{i-1}$ and the output is a stack of $L$ vectors $\mathbf{a}_i^{(l)}$ $(l=1,2,...,L)$ of length $n_i$. The operation that takes the stack of $K$ input vectors to the $l^{th}$ output vector is called a \textit{feature map}, and is defined by the operation:
{\ron
\begin{equation}
\mathbf{a}_i^{(l)} = \phi \left(\sum_{k=1}^{K}\mathbf{w}_i^{(k,l)} * \mathbf{a}_{i-1}^{(k)} + \mathbf{b}_i^{(l)} \right)
\end{equation}
where $*$ is the discrete cross-correlation operation (colloquially called ``convolution''), $\mathbf{w}_i^{(k,l)}$ is a vector of length $m_i$ of learned parameters called the convolution \textit{kernel} or \textit{filter}, $\mathbf{b}_i^{(l)}$ is a vector of length $n_i$ of learned bias parameters}}, and $\phi$ is an elementwise activation function. Typically the kernel size is small (e.g. {\ron $m_i=3$ or 5}) and the feature map detects the presence of a local feature along its input.

A pooling layer aggregates values within small neighborhoods along its input, typically by taking the mean or maximum value within each neighborhood. This makes the network approximately invariant to small translations of the input, which is helpful if we care more about whether a particular feature is present than its precise location. A pooling layer typically aggregates regions spaced $s$ neurons apart, rather than one neuron apart, where $s$ is called the \textit{stride length} (see Figure \ref{cnn}). This reduces the number of neurons in the pooling layer, which allows the feature maps in the next convolutional layer to have a wider view of the input -- albeit at a lower resolution -- without increasing the number of trainable parameters.

\subsection{Neural Network Training}\label{nntraining}

A neural network is trained to minimize a \textit{cost function}, which is a measure of how far its predictions are from the true labels in its training set. The cross-entropy error function is the standard cost function for binary classification, and is defined by the equation:

\begin{equation}\label{eq:crossentropy}
    C( \mathbf{w} ) = - \frac{1}{M} \sum_{i=1}^{M} \left( y_i \log \hat{y}_i + (1-y_i) \log (1 - \hat{y}_i) \right)
\end{equation}

where $\mathbf{w}$ is the vector of all parameters in the model, $y_1, y_2, ..., y_M$ are the true labels of all examples in the training set (defined to be either 0 or 1, depending on their classifications), and $\hat{y}_i$ is the model's predicted probability that $y_i=1$ given $\mathbf{w}$.

The gradient of the cost function with respect to the model's parameters indicates how those parameters should be changed to reduce the value of the cost function. The model's parameters, which are initially random, are iteratively updated by descending along the gradient until a suitable minimum value of the cost function is reached. It is unnecessary and computationally expensive to compute the exact gradient of the cost function (e.g. computing the exact gradient of $C( \mathbf{w} )$ in \eqref{eq:crossentropy} would involve iterating over the entire training set). Instead, each gradient step approximates the true gradient using a subset of the training set (called a ``mini-batch'') of size $B$, where $1 \le B \ll M$. $B$ is typically called the ``(mini-)batch size.'' 
%Another important hyperparameter (that is, a pre-defined parameter not learned by the model) is the ``learning rate'', which controls the magnitude of the parameter updates at each gradient step.

% The backpropagation algorithm is used to efficiently compute the gradient of the cost function with respect to the parameters of the neural network, which are initially random. The gradient indicates how the parameters should be changed in order to reduce the value of the cost function. This process is repeated over batches of training examples until the model is considered to have converged. A common practice is to train the model for a fixed number of epochs -- that is, to process the entire data set a fixed number of times.

\section{Creating our Training Set}\label{trainingset}

\subsection{TCEs and Labels}\label{tceslabels}

We derived our training set of labeled TCEs from the Autovetter Planet Candidate Catalog for Q1-Q17 DR24 \citep{autovettercat}, which is hosted at the NASA Exoplanet Archive\footnote{http://exoplanetarchive.ipac.caltech.edu/}.

We sourced TCE labels from the catalog's \textbf{av\_training\_set} column, which has three possible values: planet candidate (PC), astrophysical false positive (AFP) and non-transiting phenomenon (NTP). We ignored TCEs with the ``unknown'' label (UNK). These labels were produced by manual vetting and other diagnostics; see \cite{autovettercat} for full details. In total there are 3,600 PCs, 9,596 AFPs and 2,541 NTPs. We binarized the labels as ``planet'' (PC) and ``not planet'' (AFP / NTP), but our model would only require minor changes to perform multi-class classification instead.

For the rest of this paper we assume that these labels are the ground truth. However, this is not totally accurate -- we discovered several mislabeled examples by manually examining a small subset of the full training set. For example, Kepler Objects of Interest (KOI) 2950.01 and 5308.01 are eclipsing binary false positives, and are labeled as such in the latest KOI table \citep{koi8}, but both are labeled PC in the Autovetter Catalog. We assume that the effects of mislabeling on our model and performance metrics are relatively minor.

We randomly partitioned our data into three subsets: training (80\%), validation (10\%) and test (10\%). We used the validation set during development to choose model hyperparameters (Section \ref{training}), and we used the test set to evaluate final model performance (Section \ref{holdoutresults}). This is the gold-standard way to evaluate model performance because the test set contains no data used to optimize the model in any way.

% This is standard practice in statistical modeling and the gold-standard way to ensure that final performance metrics are not computed using data used to optimize the model in any way.

\subsection{Light Curves}\label{lightcurves}

We downloaded light curves produced by the \Kepler\ mission from the Mikulski Archive for Space Telescopes\footnote{http://archive.stsci.edu/}. These light curves were produced by the \Kepler\ pipeline \citep{jenkins-overview}, which calibrates the pixels, identifies optimal photometric apertures, performs simple (stationary) aperture photometry, and removes common-mode instrumental artifacts using the Presearch Data Conditioning--Maximum A Posteriori (PDCMAP) method \citep{stumpe, smithpdc, stumpe2}. Each light curve consists of integrated flux measurements spaced at 29.4 minute intervals for up to 4 years (approximately 70,000 points).

We then performed a few additional steps to prepare the light curves to be used as inputs to our neural network. For each TCE in the training set, we removed points corresponding to transits of any other confirmed planets in the system. Then we ``flattened'' the light curve (that is, we removed low frequency variability) by fitting a basis spline to the light curve and dividing the light curve by the best-fit spline. To preserve transits, we removed the TCE's in-transit points while fitting the spline and linearly interpolated over those transits. We iteratively fit the spline, removed $3 \sigma$ outliers, and re-fit the spline while interpolating over those outliers to prevent the spline from being ``pulled'' by discrepant points like cosmic ray hits. This process is illustrated in Figure 3 of \citet{vj14}.
% \footnote{This was intended to remove the effect of other transiting planets, especially when searching for new planets in known multi-planet systems (Section \ref{newcandidates}).}

One parameter we had to choose when fitting the spline was the spacing of spline ``knots'' or ``break-points'', between which the spline calculates its piecewise polynomial coefficients. Each \Kepler\ light curve has different low-frequency stellar variability characteristics, so using one break-point spacing for all light curves is sub-optimal. Instead, we chose the optimal spacing of spline break-points for each light curve by fitting splines with different break-point spacings, calculating the Bayesian Information Criterion \citep[BIC,][]{schwarz} for each spline, and choosing the break-point spacing that minimized the BIC.

\subsection{Input Representations}\label{inputrep}

\begin{figure*}[t!]
\centering
\includegraphics[width=\textwidth]{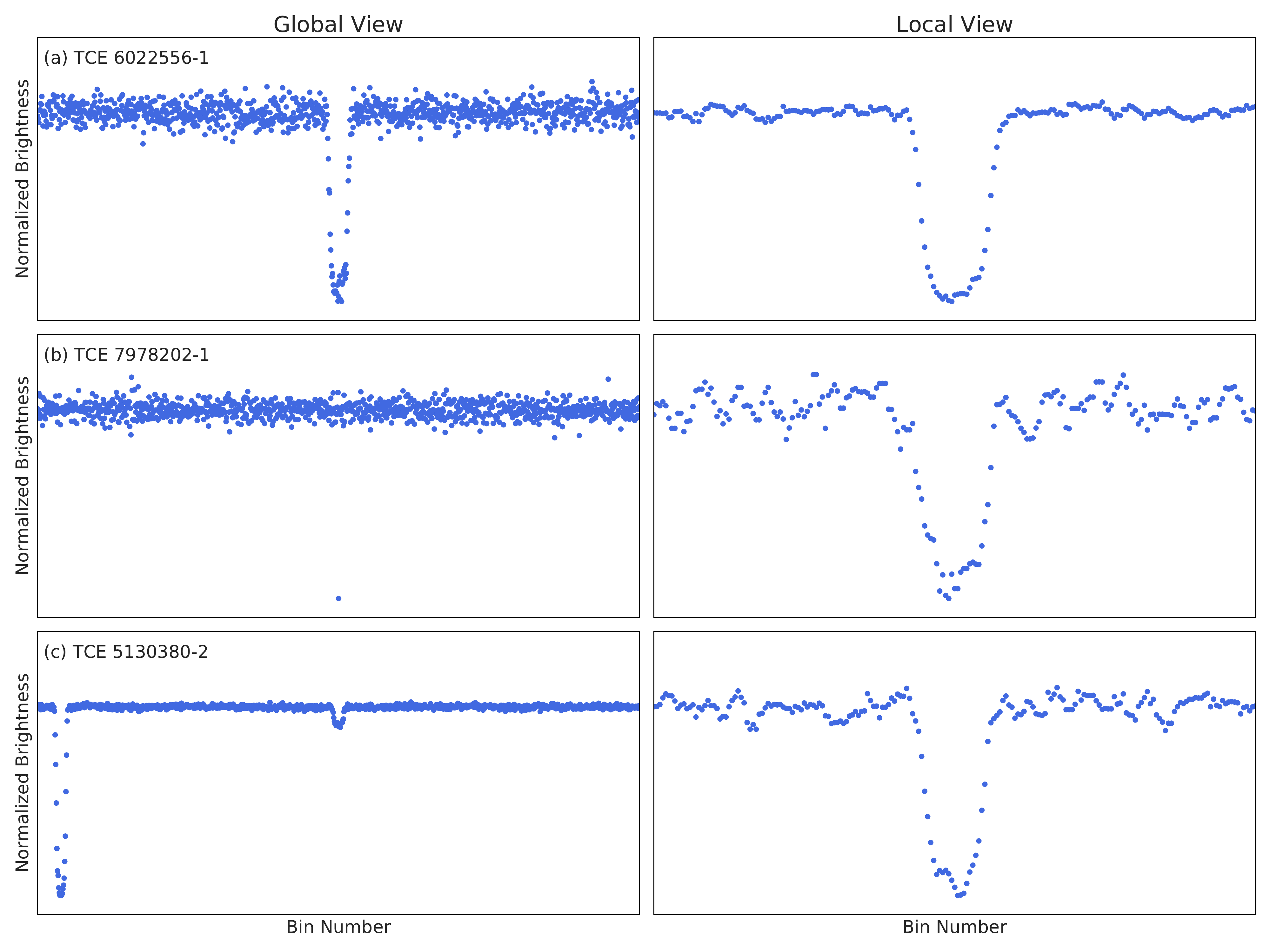}
\caption{Light curve representations that we use as inputs to our neural network models. The ``global view'' is a fixed-length representation of the entire light curve, and the ``local view'' is a fixed-length representation of a window around the detected transit. (a) Strong planet candidate. (b) Long-period planet where the transit falls into just one bin in the global view. (c) Secondary eclipse that looks like a planet in the local view. } \label{localglobal}
\end{figure*}

We generated our neural network inputs by folding each flattened light curve on the TCE period (with the event centered) and binning to produce a 1D vector.

To bin a folded light curve, we define a sequence of uniform intervals on the time axis with width $\delta$ and distance $\lambda$ between midpoints, and compute the median flux of the points falling within each interval. If we choose $\delta = \lambda$ then the intervals partition the time axis: each point falls in precisely one bin. If we choose $\delta > \lambda$ then the bins overlap, which reduces scatter and makes some transits more visible.

The simplest choice of $\lambda$ would be a fixed constant, but this would result in a wide range of input lengths due to the wide range of TCE periods. Instead we choose two TCE-specific values for $\lambda$, each of which generates a different ``view'' of the light curve (Figure \ref{localglobal}).

We generate a \textit{global view} of the light curve by choosing $\lambda$ as a fraction of the TCE period. All light curves are binned to the same length and each bin represents the same number of points, on average, across light curves. A disadvantage is that long-period TCEs may end up with very narrow transits that fall entirely within a small number of bins (Figure \ref{localglobal}b).

We generate a \textit{local view} of the transit by choosing $\lambda$ as a fraction of the TCE duration. We consider $k$ transit durations on either side of the event so that the transit occupies a fixed fraction of the resulting vector. This technique represents short- and long-period TCEs equally, but it only looks at part of the curve and therefore may miss important information, such as secondary eclipses (Figure \ref{localglobal}c)

Similar techniques are used by \citet{armstrong} (local binning) and \citet{lpp} (global binning away from the transit and local binning near the transit). Unlike those papers, we use these two representations as separate inputs to our model.

Finally, we normalized all light curves to have median 0 and minimum value -1 so that all TCEs had a fixed transit depth.

\section{Automatic Vetting Models}
\label{automaticvetting}
\subsection{Neural Network Architectures}

We consider 3 types of neural network for classifying \Kepler\ TCEs as either ``planets'' or ``not planets''. For each type, we consider 3 different input options: just the global view, just the local view, and both the global and local views.

\begin{figure}[t!]
\centering
\includegraphics[width=0.8\columnwidth]{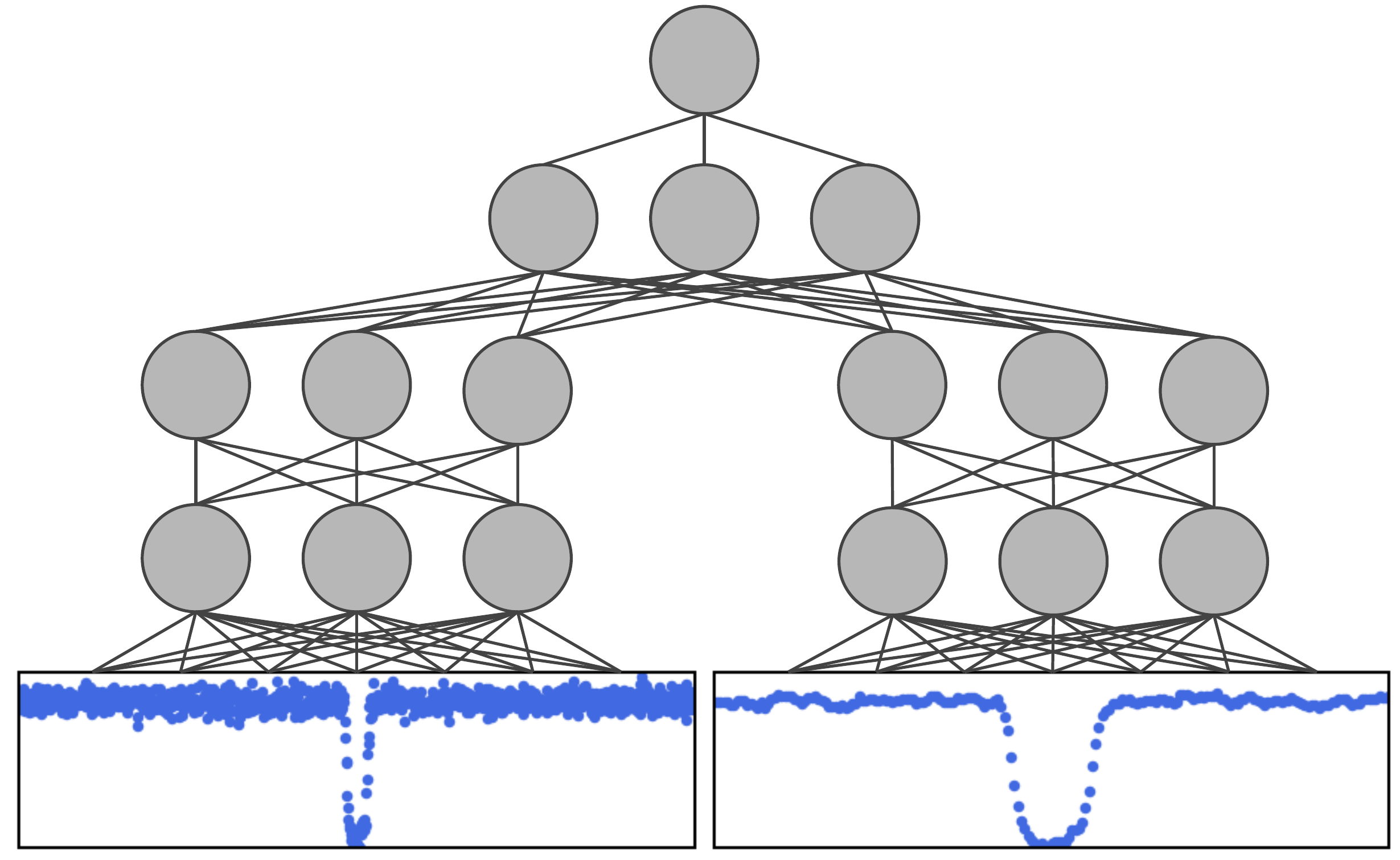}
\caption{Fully connected neural network architecture for classifying light curves, with both global and local input views.} \label{fcmodel}
\end{figure}

\begin{figure}[t!]
\centering
\includegraphics[width=0.8\columnwidth]{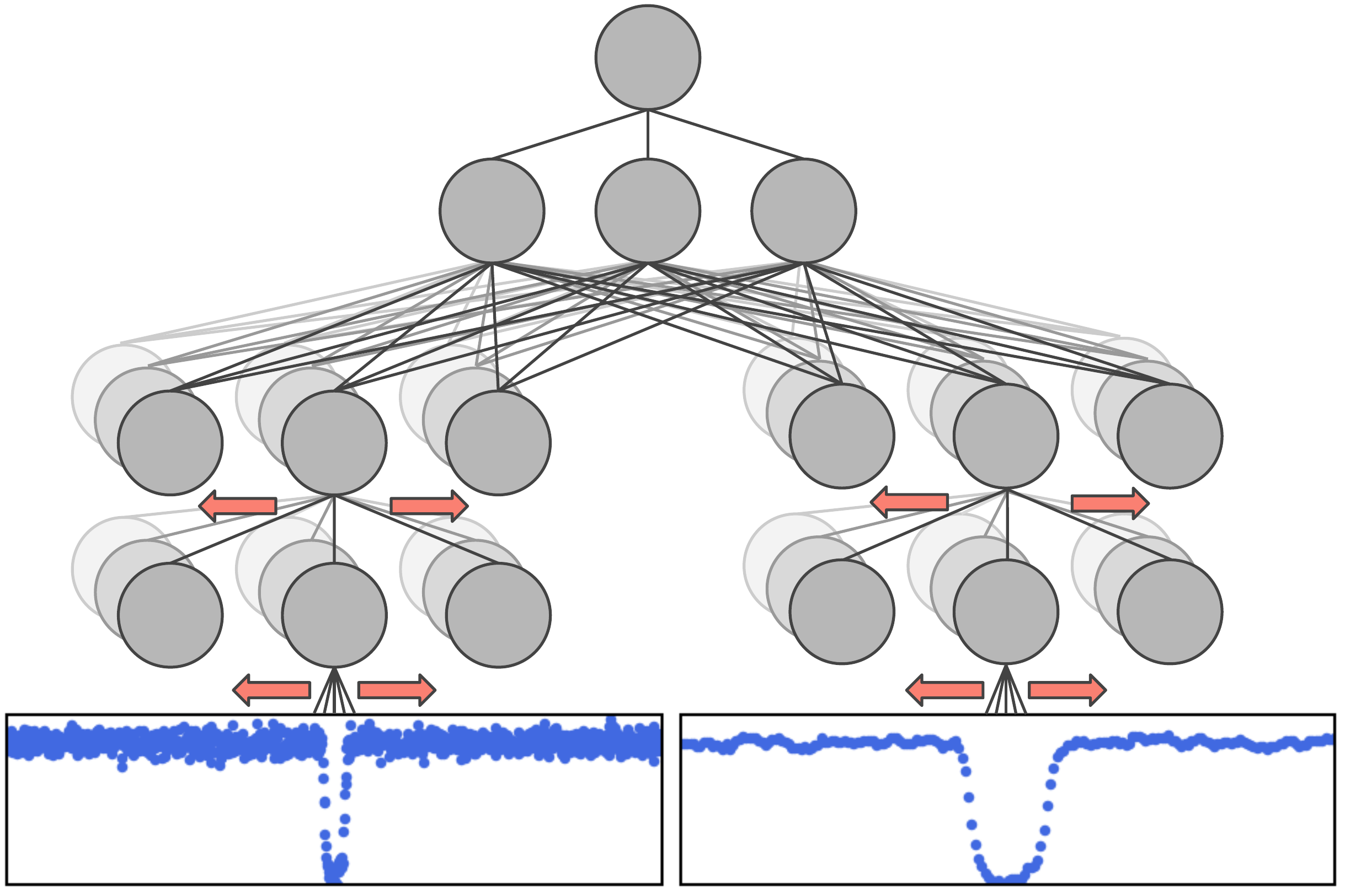}
\caption{Convolutional neural network architecture for classifying light curves, with both global and local input views.} \label{cnnmodel}
\end{figure}

\begin{itemize}
    \item \textbf{Linear architecture.} Our baseline model is a neural network with zero hidden layers (which is equivalent to a linear logistic regression model). This is a natural choice for its simplicity and popularity, but it makes the strong assumption that the input data is linearly separable -- that is, planets and non-planets are separated by a linear decision surface in the input space. If both global and local views are present, we concatenate them into a single input vector.
    % The natural choice of baseline is the simplest possible neural network -- one with zero hidden layers. This is equivalent to linear logistic regression, which is simple and widely known and used, but it makes the strong assumption that the input data is linearly separable -- that is, planets can be separated exactly from non-planets by a linear decision surface in the input space.
    \item \textbf{Fully connected architecture.} A fully connected neural network is the most general type of neural network and makes the fewest assumptions about the input data. If both global and local views are present, we pass the two vectors through disjoint columns of fully connected layers before combining them in shared fully connected layers (Figure \ref{fcmodel}).
    \item \textbf{Convolutional architecture.} Convolutional neural networks have been tremendously successful in applications with spatially structured input data, including speech synthesis \citep{wavenet} and image classification \citep{alexnet}. We use a 1-dimensional convolutional neural network with max pooling. This architecture assumes that input light curves can be described by spatially local features, and that the output of the network should be invariant to small translations of the input. If both global and local views are present, we pass the two vectors through disjoint convolutional columns before combining them in shared fully connected layers (Figure \ref{cnnmodel}).
\end{itemize}

All hidden layers use the ReLU (linear rectifier) activation function, and the output layer uses the sigmoid activation function. The output of each model is the predicted probability that the input is a transiting planet; values close to 1 indicate high confidence that the input is a transiting planet and values close to 0 indicate high confidence that the input is a false positive.

\subsection{Training Procedure} \label{training}

We implemented our models in TensorFlow, an open source software library for numerical computation and machine learning \citep{tensorflow}. We used the Adam optimization algorithm \citep{adam} to minimize the cross-entropy error function over the training set (Section \ref{nntraining}). We augmented our training data by applying random horizontal reflections to the light curves during training. We also applied dropout regularization to the fully connected layers, which helps prevent overfitting by randomly ``dropping'' some of the output neurons from each layer during training to prevent the model becoming overly reliant on any of its features \citep{dropout}.

We used the Google-Vizier system for black-box optimization \citep{vizier} to automatically tune our hyperparameters, including those for the input representations (e.g. number of bins, bin width), model architecture (e.g. number of fully connected layers, number of convolutional layers, kernel size), and training (e.g. dropout probability). Each Vizier ``study'' trained several thousand models to find the hyperparameter configuration that maximized {\ron the area under the receiver operating characteristic curve (see Section \ref{holdoutresults})} over the validation set. {\ron Each model was trained on a single central processing unit (CPU), and training time ranged from 10 minutes to 5 hours depending on the size of the model. Our best model (see Section \ref{bestmodel}) took 90 minutes to train. To speed up the hyperparameter search, we used 100 CPUs per study to train individual models in parallel.} We ran many Vizier studies during model development as we iteratively improved our input representations and design decisions. We tuned each combination of architecture and input representation separately.

\subsection{Model Averaging}

Once we had selected the optimal hyperparameters for a particular architecture and input representation, we trained 10 independent copies with different random parameter initializations. We used the average outputs of these 10 copies for all predictions in Sections \ref{analysis} and \ref{newcandidates}. This technique, known as ``model averaging'' (a type of ``ensembling''), often improves performance because different versions of the same configuration may perform better or worse on different regions of the input space, especially when the training set is small and overfitting is more likely. It also makes different configurations more comparable by reducing the variance that exists between individual models.

\section{Model Analysis}\label{analysis}

\subsection{Test Set Performance}\label{holdoutresults}

{\ron
We use the following metrics to assess our model's performance:

\begin{itemize}
    \item \textbf{Precision}: the fraction of signals classified as planets that are true planets \citep[also known as reliability, see, e.g.][]{koi8}.
    \item \textbf{Recall}: the fraction of true planets that are classified as planets (also known as completeness).
    \item \textbf{Accuracy}: the fraction of correct classifications.
    \item \textbf{AUC}: the Area Under the receiver operating characteristic Curve, which is equivalent to the probability that a randomly selected planet is scored higher than a randomly selected false positive.
\end{itemize}

The values of the first three metrics depend on the classification threshold we choose for our models.} A natural choice is to classify a TCE as a planet candidate if its predicted probability of being a planet is above 0.5, and as a false positive if its predicted probability is below 0.5, but this threshold can be adjusted to trade-off {\ron precision} versus {\ron recall}. Increasing the threshold typically results in higher {\ron precision} at the expense of lower {\ron recall}, and vice-versa. {\ron The value of AUC is independent of the choice of classification threshold.}

\begin{figure}[t!]
\centering
\includegraphics[width=\columnwidth]{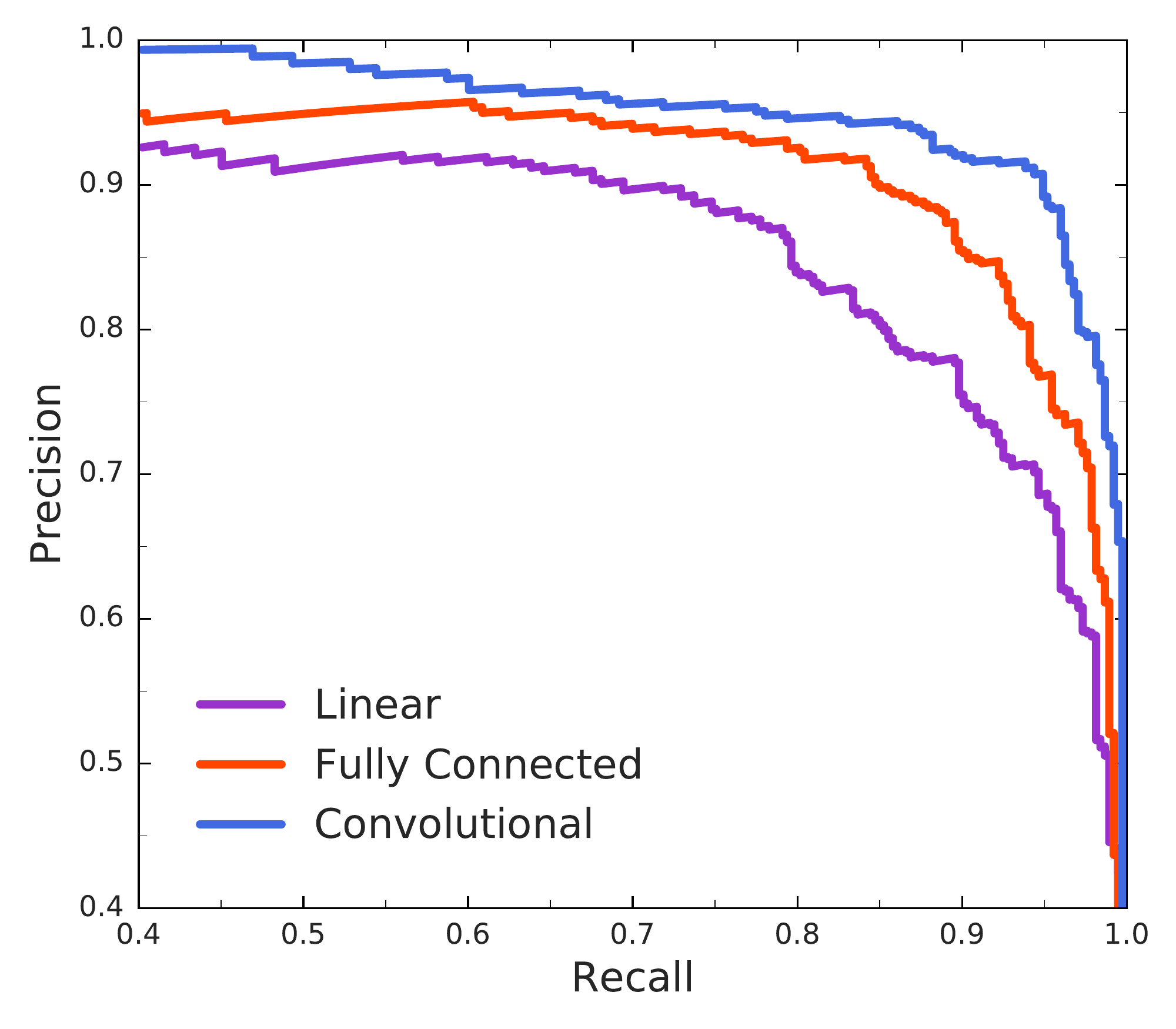}
\caption{{\ron Precision versus recall} on the test set for our three neural network architectures. All models use both global and local input views. For each model, every point on its curve corresponds to a different choice of classification threshold. All TCEs predicted to be planets with probability greater than the threshold are considered to be classified as planets, and all TCEs predicted below the threshold are considered to be classified as false positives. Our convolutional architecture has the best overall performance. For example, with an appropriate choice of threshold, it has {\ron recall} of 0.95 (that is, 95\% of real planets are classified as planets) at {\ron precision} of 0.90 (that is, 90\% of its classifications are real planets).} \label{precisionrecall}
\end{figure}

Figure \ref{precisionrecall} shows {\ron precision} versus {\ron recall} for our three model architectures on our test set, which consists of 1,523 TCEs that were not used to train our models or inform our hyperparameter decisions. All models in Figure \ref{precisionrecall} use both global and local input views. Our convolutional architecture performs the best, followed by our fully connected architecture. A full description of our best model's configuration is presented in Section \ref{bestmodel}.

\begin{deluxetable}{lccc}[t!]
\tablewidth{250pt}
\tablecaption{Test Set Accuracy \label{accuracy}}
\tablehead{
\colhead{} & \colhead{Global} &\colhead{Local} &  \colhead{Global \& Local}}
\startdata
Linear           & 0.869  & 0.879  & 0.917  \\
Fully Connected  & 0.902  & 0.912  & 0.941  \\
Convolutional    & 0.954  & 0.924  & 0.960  \\
\enddata
\tablenotetext{}{The fraction of correct classifications on the test set for each choice of neural network architecture and input representation. For each model, all TCEs predicted to be planets with probability greater than 0.5 are considered to be classified as planets, and all TCEs predicted below 0.5 are considered to be classified as false positives.}
\end{deluxetable}

\begin{deluxetable}{lccc}[t!]
\tablewidth{250pt}
\tablecaption{Test Set AUC \label{auc}}
\tablehead{
\colhead{} & \colhead{Global} &\colhead{Local} &  \colhead{Global \& Local}}
\startdata
Linear           & 0.922  & 0.933  & 0.963  \\
Fully Connected  & 0.961  & 0.965  & 0.977  \\
Convolutional    & 0.985  & 0.973  & 0.988  \\
\enddata
\tablenotetext{}{AUC on the test set for each choice of neural network architecture and input representation. AUC is the probability that a randomly selected planet candidate is predicted to be a planet with higher probability than a randomly selected false positive. It {\ron measures a model's ability to rank TCEs: the maximum AUC value of 1 would indicate that all planets are ranked higher than all false positives.}}
\end{deluxetable}

{\ron Tables \ref{accuracy}-\ref{auc} show the accuracy and AUC, respectively, of each combination of model architecture and input representation on our test set. These tables} show that, for all architectures, using both global and local views gives better performance than using either view individually. Interestingly, the local view (on its own) gives better performance than the global view (on its own) in the linear and fully connected architectures, but the opposite is true in the convolutional architecture. We observed that the linear and fully connected architectures have difficulty distinguishing U-shaped transits (like planets) from V-shaped transits (like eclipsing binaries) when the transit width is not normalized in the global view, but the convolutional architecture is able to distinguish U-shaped transits from V-shaped transits at different widths, and is also better at identifying secondary eclipses in the global view.

\subsection{Best Model Configuration} \label{bestmodel}

\begin{figure}[t!]
\centering
\includegraphics[width=0.8\columnwidth]{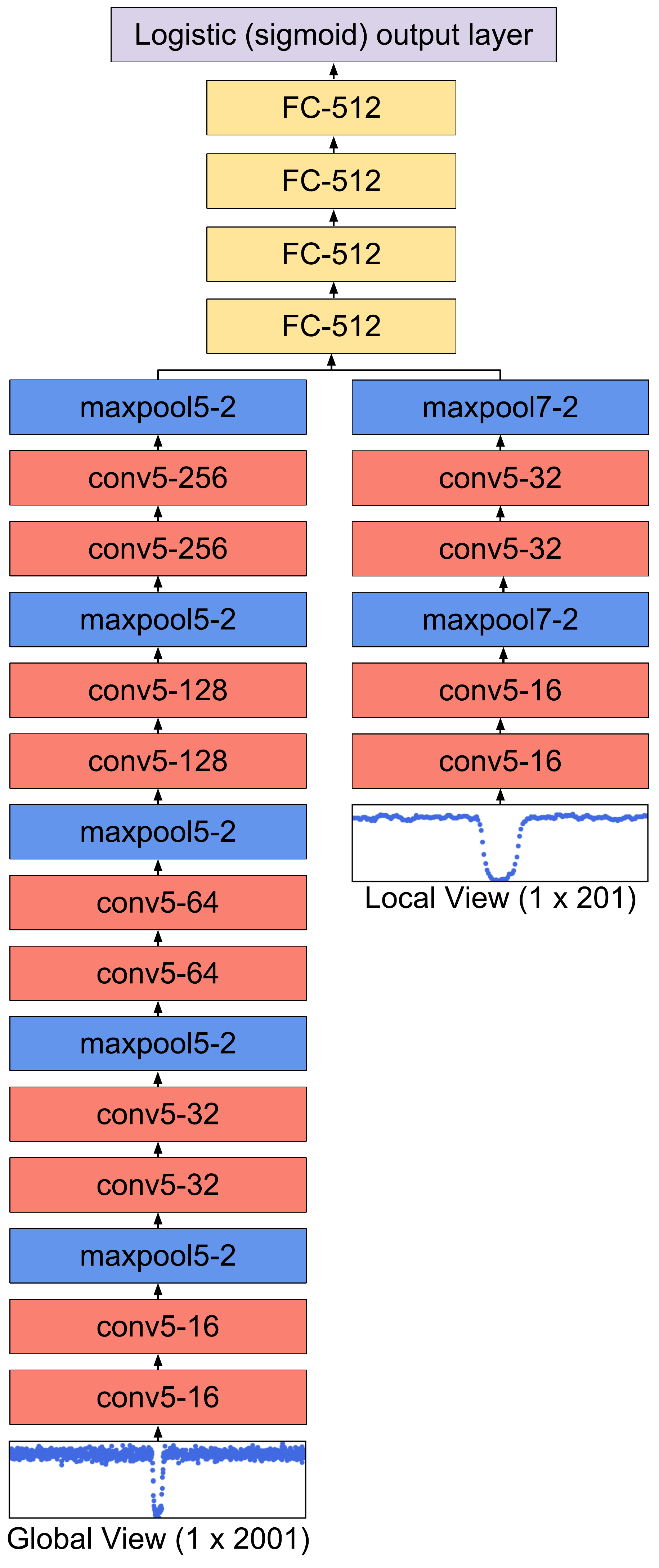}
\caption{The architecture of our best performing neural network model. Convolutional layers are denoted conv\textlangle kernel size\textrangle -\textlangle number of feature maps\textrangle, max pooling layers are denoted maxpool\textlangle window length\textrangle -\textlangle stride length\textrangle , and fully connected layers are denoted FC-\textlangle number of units\textrangle .} \label{fig:bestarchitecture}
\end{figure}

Our best model, based on test set performance, uses the convolutional architecture with global and local input views. Its exact configuration is presented in Figure \ref{fig:bestarchitecture}. Its hyperparameters were chosen to maximize AUC on the validation set using the procedure described in Section \ref{training}. Its performance on the test set is presented in Section \ref{holdoutresults}.

We used a global view of length 2001 with $\lambda = \delta = p / 2001$ (where $p$ is the TCE period), and a local view of length 201 with $k=4$ transit durations on either size of the event, $\lambda = 2kd / 201$, and $\delta = 0.16d$ (where $d$ is the TCE duration). See Section \ref{inputrep} for a description of these parameters.

We trained the model using a batch size of 64 for 50 epochs. We used the Adam optimization algorithm with $\alpha = 10^{-5}$, $\beta_1 = 0.9$, $\beta_2 = 0.999$, and $\epsilon = 10^{-8}$ (see \citealt{adam} for a description of these parameters). We did not use dropout regularization in this configuration.

\subsection{Visualizing the Model}\label{visualizing}

\subsubsection{Prediction Heatmaps}\label{heatmaps}

One way to visualize a convolutional neural network is to systematically occlude portions of the input to see how the output prediction changes \citep{understandingcnn}. We can use this technique to create heat maps showing the most important regions of the input contributing to a classification decision.

\begin{figure}[t!]
%\vspace{1mm}  % This prevents a single line of text appearing under the figure.
\centering
\includegraphics[width=\columnwidth]{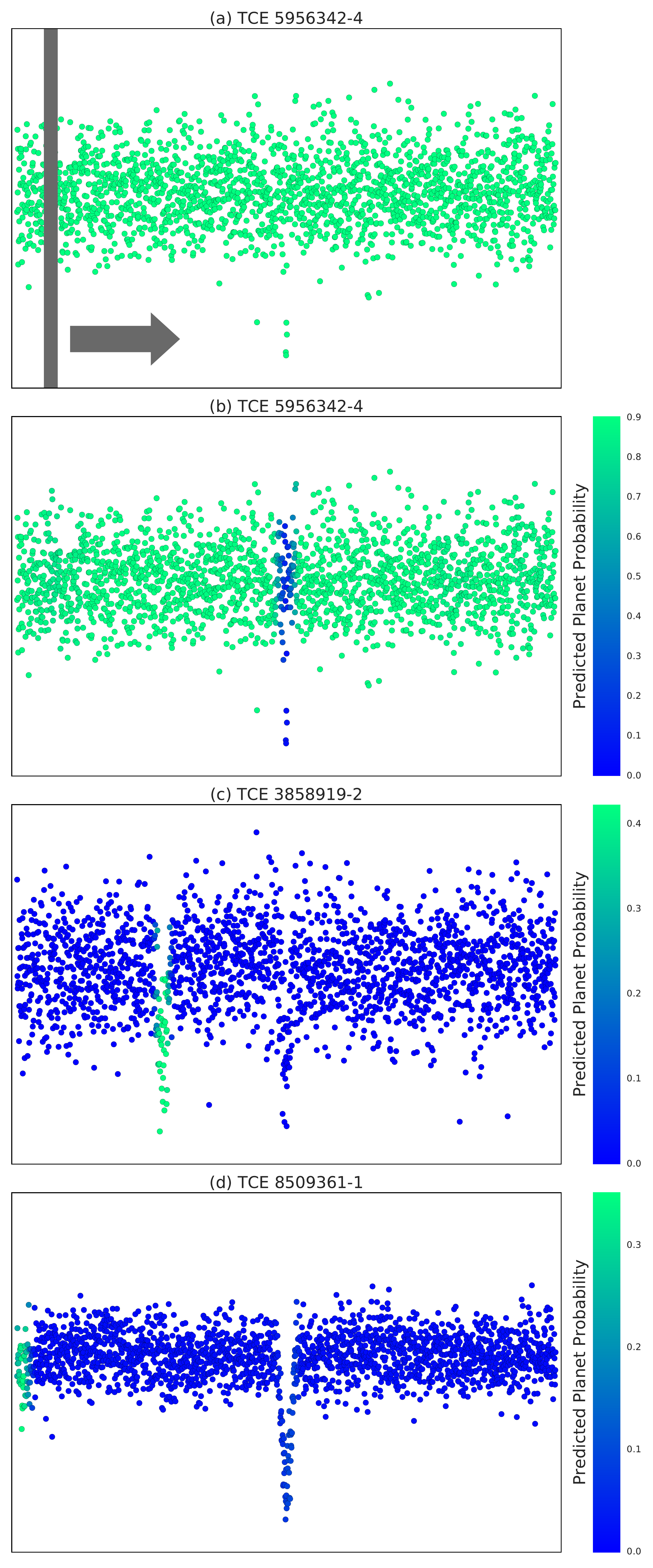}
\caption{Visualizing the model's focus. Regions that are particularly important to the model's decision change the output prediction when occluded. (a) We occluded points in a sliding window of width 50 points (shown to scale). {\ron The color scale for each point shows the mean predicted probability of being a planet over all predictions where that point was occluded.} (b) Transiting planet: the model's prediction drops to zero when the transit is occluded. (c,d) Eclipsing binaries: the model identifies secondary eclipses, even in subtle cases like (d). } \label{cnnheatmaps}
\end{figure}

Figure \ref{cnnheatmaps} shows prediction heat maps of a convolutional model using the global view. We occluded points in a sliding window of width 50  points by setting their values to zero (Figure \ref{cnnheatmaps}a). We computed the heat map value for each point by averaging the model's {\ron predicted probability of being a planet} whenever that point was occluded.

Figure \ref{cnnheatmaps}b shows that the transit is the most important region of the input for a planet candidate. The model's prediction drops to zero when the transit is occluded, but is largely unchanged when any other region is blocked.

Figure \ref{cnnheatmaps}c,d show that the model learns to identify secondary eclipses. The model's planet prediction increases when a secondary eclipse is occluded because we are hiding the evidence that this TCE is an eclipsing binary. Note that even when the secondary eclipse is blocked, the model's predictions are still low ($\lesssim 0.4$) in Figure \ref{cnnheatmaps}c,d, likely because the primary transit is V-shaped, which is also evidence of an eclipsing binary.

\subsubsection{t-SNE Embeddings}

\begin{figure*}[p!]
\centering
\includegraphics[width=\textwidth]{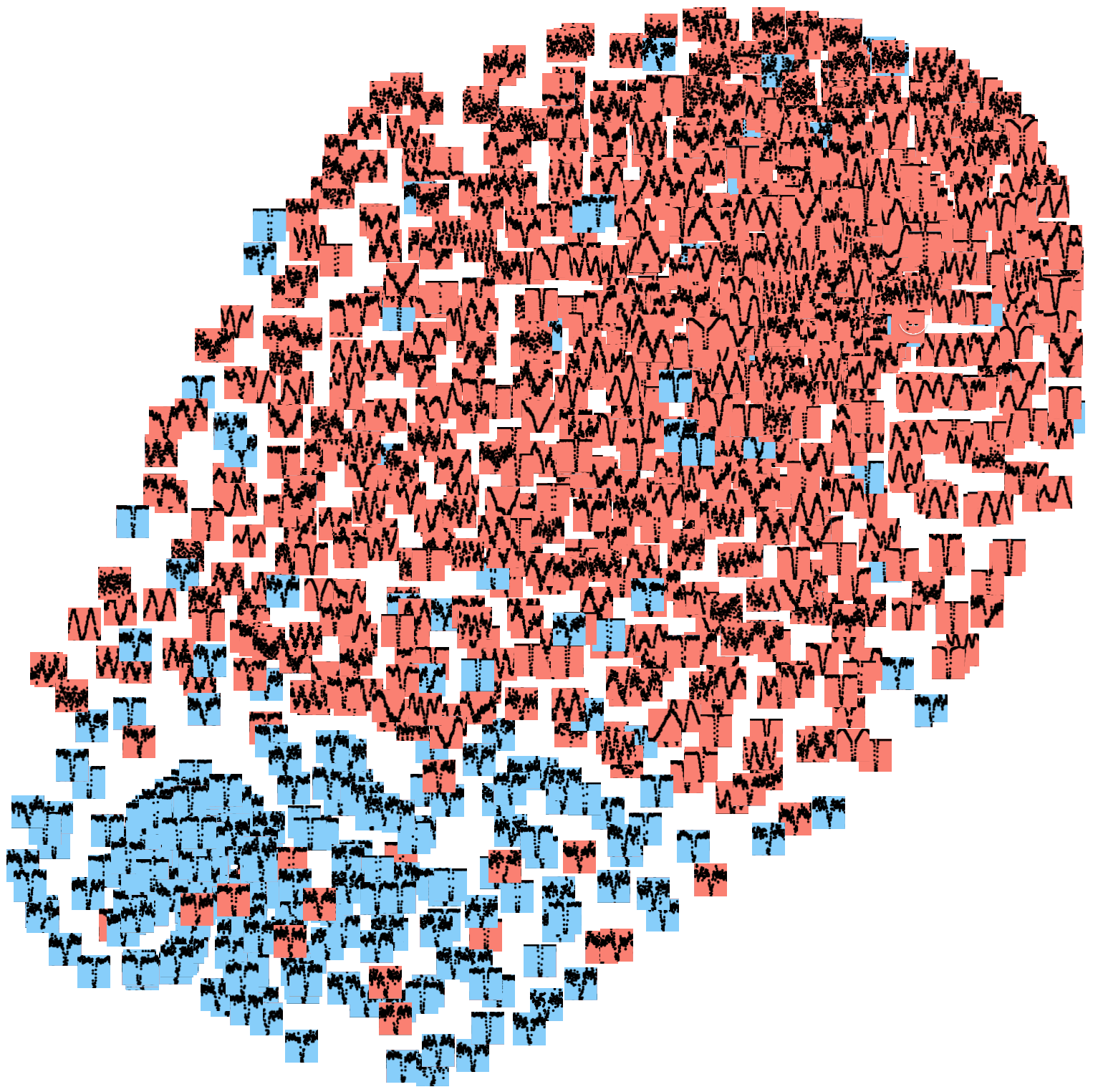}
\caption{Visualizing the geometric space in which our convolutional neural network embeds its input light curves. We used the t-SNE algorithm for dimensionality reduction to create a {\ron 2}-dimensional embedding such that light curves that are close in the original space are also close in the {\ron 2}D space. Light curves that are far apart in {\ron 2}D space are not necessarily far apart in the original space. Notice that true planet candidates (colored blue) are close to other planet candidates, and false positives (colored red) are close to similarly-shaped false positives.} \label{tsnefig}
\end{figure*}

We can consider the final hidden layer of a neural network to be a learned representation of the input such that a linear classifier can distinguish between the different classes. Geometrically, the values of the final hidden layer (which are points in $d$-dimensional space) are such that a linear boundary in $\mathbb{R}^d$ can separate planet candidates from false positives.

To visualize this space in {\ron 2} dimensions, we use t-Distributed Stochastic Neighbor Embedding \citep[t-SNE,][]{tsne}, a technique for dimensionality reduction that is well suited for visualizing high-dimensional data sets. Figure \ref{tsnefig} shows a {\ron 2}-dimensional t-SNE embedding of the final layer activations of our best model using TCEs from the test set.

\subsection{Performance on Simulated Data}\label{simdata}

The \Kepler\ team generated multiple kinds of simulated data to test the performance of the \Kepler\ pipeline (which detects TCEs in \Kepler\ light curves) and the Robovetter (which classifies TCEs by automating the manual review process developed by the TCE Review Team) \citep{dr25-kp-efficiency, dr25-rv-efficiency, koi8}. The Robovetter differs from machine learning vetting systems (like ours) in that its heuristics are explicitly defined by humans instead of learned automatically from data\footnote{The Robovetter has two heuristics based on the \textit{LPP} metric, which is computed by a machine learning model \citep[][]{lpp}, but the heuristics applied to this metric are explicitly defined.}.

The \Kepler\ team recently released their simulated data products to the public, so we decided to test the effectiveness of our model at distinguishing simulated planets from simulated false positives. We considered the following simulated data sets, which are available at the NASA Expolanet Archive:

\begin{itemize}
    \item \textbf{Injected Group 1} contains a single on-target injected periodic signal for $146,294$ \Kepler\ targets. Each injection simulates a transiting planet.
    \item \textbf{Injected Group 2} contains a single off-target injected periodic signal for $22,978$ \Kepler\ targets. Each injection simulates an off-target transiting planet or eclipsing binary false positive.
    \item \textbf{Injected Group 3} contains two on-target injected periodic signals, with a common period, for $9,856$ \Kepler\ targets. Each injection simulates an eclipsing binary false positive.
    \item \textbf{Inverted Group} contains $198,640$ \Kepler\ targets that have had their light curves inverted. Any detected transits are likely to be false positives.
\end{itemize}

The \Kepler\ pipeline uses the multiple event statistic \citep[MES,][]{jenkins-mes} to measure the strength of a transit signal relative to the noise. A detected signal must have MES \textgreater 7.1 to be considered significant enough to become a TCE. Therefore, the \Kepler\ pipeline only recovers a subset of the simulated planets and false positives as TCEs. Table \ref{simclassifications} shows the number of these simulated TCEs classified as planets by our model with probability greater than 0.5, compared to the Robovetter. We discuss these results in the following sections.

\begin{deluxetable}{lrrr}[t!]
\tablewidth{250pt}
\tablecaption{Simulated Data Classfications \label{simclassifications}}
\tablehead{
\colhead{Simulated Group} & \colhead{TCEs} & \colhead{Convolutional Model} &\colhead{Robovetter}}
\startdata
Injected 1            &  45,377  &  36,826 {\ron (81.2\%)}  &  38,668 {\ron (85.2\%)}  \\
Injected 2            &  18,897  &  17,044 {\ron (90.2\%)}  &  8,805 {\ron (46.6\%)}   \\
Injected 3            &  10,502  &  6,852 {\ron (65.2\%)}   &  3,932 {\ron (37.4\%)}   \\
Inverted              &  19,536  &  621 {\ron (3.18\%)}     &  95 {\ron (0.49\%)}      \\
Inverted (cleaned)    &  14,953  &  539 {\ron (3.60\%)}    &  75 {\ron (0.50\%)}      \\
\enddata
%\tablenotetext{1}{Number of TCEs scored above 0.5.}
%\tablenotetext{2}{Number of TCEs classified as Planet Candidates.}
\tablenotetext{}{The number of simulated TCEs predicted to be planets by our convolutional neural network (exceeding the probability threshold of 0.5) and the Robovetter (classified as Planet Candidates). Injected Group 1 are simulated planets, Injected Group 2 are simulated background planets or eclipsing binaries, Injected Group 3 are simulated eclipsing binaries, and the Inverted Group are false positives from inverted light curves. We also include a ``cleaned'' subset of the Inverted Group, from which certain types of signals known to look like inverted transits have been removed \citep{koi8}.}
\end{deluxetable}

\subsubsection{Injected Group 1}

\begin{figure}[t!]
\centering
\includegraphics[width=\columnwidth]{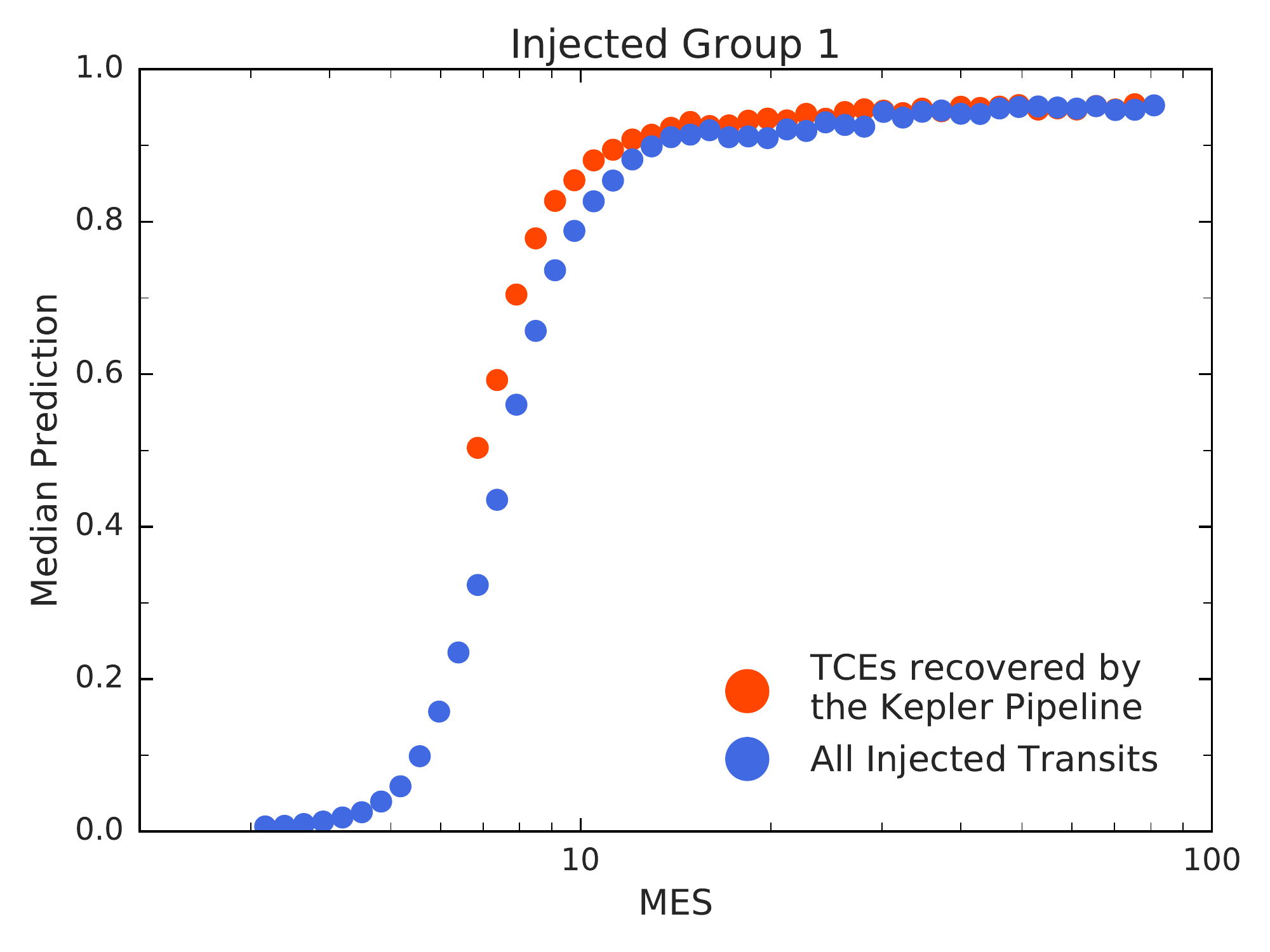}
\caption{The median predicted probability of being a planet by our model, versus MES, for all simulated planet transits in Injected Group 1. The red points show TCEs recovered by the \Kepler\ pipeline, at the detected period and epoch, versus MES measured by the  \Kepler\ pipeline. The blue points show injected TCEs, at the injected period and epoch, versus ``expected MES'', which was the intended value of MES when injecting the signal. Note that measured MES is systematically slightly lower than expected MES \citep{dr25-kp-efficiency}. \label{inj1}}
\end{figure}

Among the 45,377 simulated planets recovered by the \Kepler\ pipeline, our model is similarly permissive to the latest version of the Robovetter, correctly classifying 36,826 of these TCEs as planets versus the Robovetter's 38,668. Among the full set of 146,294 TCEs (many of which were not recovered by the \Kepler\ pipeline), we find that our model's predictions depend strongly on the MES of the simulated signal (see Figure \ref{inj1}). At MES greater than 10, our model typically has high confidence that injected planet signals are real, but our model's confidence declines steeply from TCEs with MES=10 to TCEs with MES=5. The region between MES=5 and MES=10 is of particular interest because: a) it is the likely region of Earth-sized planets transiting Sun-like stars in temperate orbits, and b) it is at the detection limit of the \Kepler\ pipeline, where the recovery rate is low and where there are likely to be as-yet undiscovered planets \citep{christiansen3,dr25-kp-efficiency}.

We suspect that we may be able to improve our model's performance on low signal-to-noise TCEs by including simulated transits like these in our training set. In our current implementation, only 8\% of planets and planet candidates in our training set have MES less than 10; our model rarely saw any low-MES positive examples during training and may have learned that low-MES events are {\em a  priori} unlikely to be true planets. Including simulated low-MES planet candidates (as well as simulated low-MES false positives) in our training set may make our model more discriminating at low MES.

%For Injected Group 1, we also computed our model's predictions on the entire set of 146,294 simulated transits (assuming the injected period and epoch were identified exactly), of which 64,826 had predictions above 0.5.
%At MES less than 5, the simulated signal is typically indistinguishable from noise, and our model rejects most TCEs.

\subsubsection{Injected Group 2}\label{inj2}

% We also ran our model on the simulated false positive signals

% We find that while our model recovers most simulated true positives at a good rate, it is not yet able to identify simulated false positives as well as the Robovetter.

We find that our model classifies many simulated false positive TCEs from Injected Group 2 as planets. The signals in Injected Group 2 were injected into the \Kepler\ pixels off-target, simulating an astrophysical signal from a nearby star, which can mimic the transit of an on-target planet\footnote{We note that in many cases, the off-target signal could be caused by a transiting planet around a background star, but these signals are still typically classified as false positives.}. The fact that our model often classifies these signals as planets makes sense, because our model does not take in any information about the location of the transits on the pixels, like centroid shifts during transit or in-transit difference imaging. These types of diagnostics can help identify when transit signals are coming from off-target sources. We have plans to incorporate this type of centroid or positional information into our model in the future to help identify this kind of false positive.

\subsubsection{Injected Group 3}

\begin{figure}[t!]
\centering
\includegraphics[width=\columnwidth]{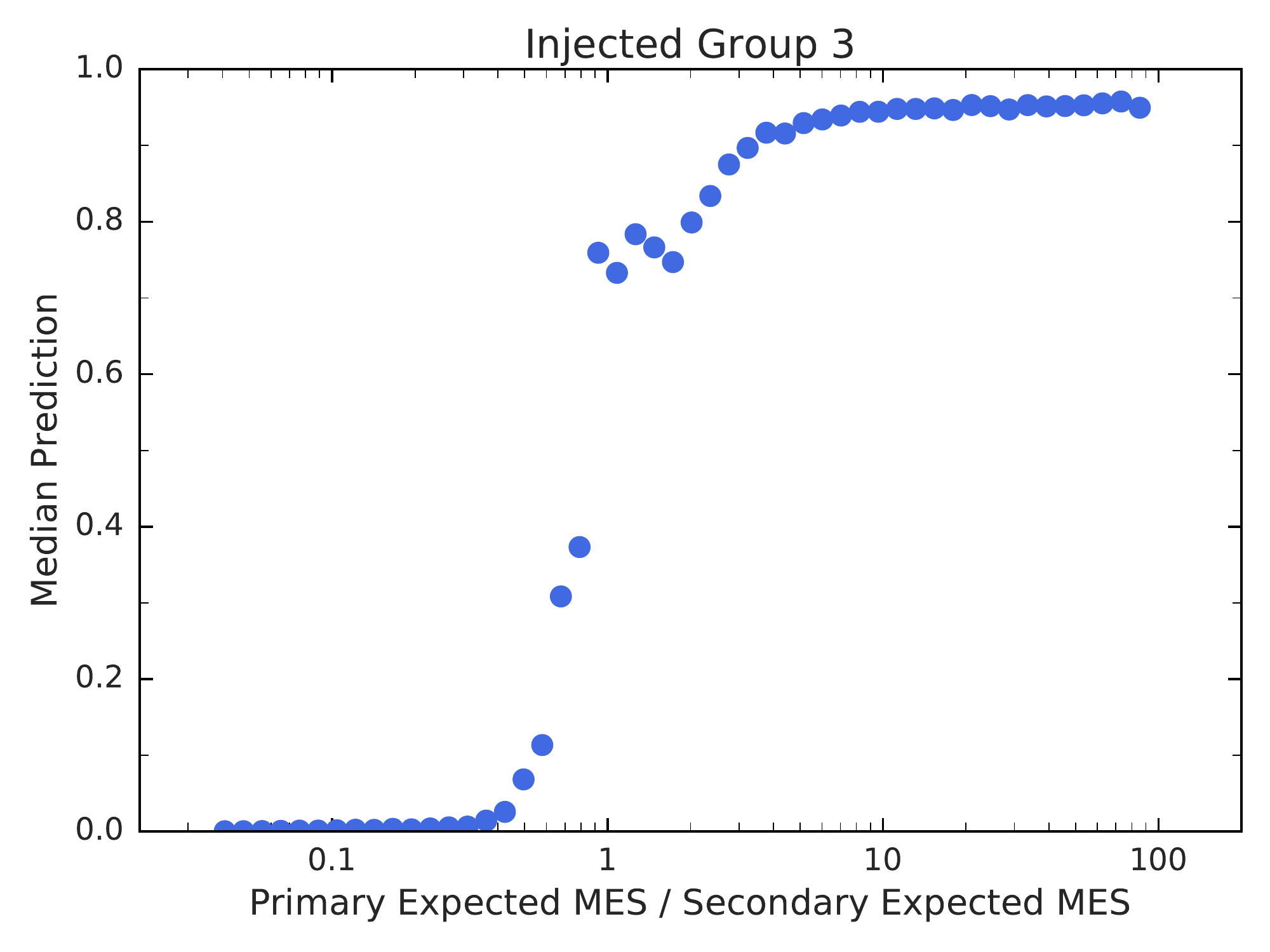}
\caption{The median predicted probability of being a planet by our model, versus primary-to-secondary expected MES ratio, for simulated eclipsing binaries in Injected Group 3 recovered by the \Kepler\ pipeline.} \label{inj3}
\end{figure}

We find that our model predicts about 65\% of the simulated eclipsing binary TCEs (with secondary events injected at the same orbital period) to be planets with probability greater than 0.5. Our model is therefore not able to identify these false positives as well as the Robovetter, which only classifies about 37\% of these simulated eclipsing binaries as planets.  Figure \ref{inj3} shows that our model rejects most simulated TCEs when the secondary signal is stronger than the primary signal, but passes most simulated TCEs when the secondary signal is weaker than the primary signal. We know from Section \ref{heatmaps} that our model can, in some cases at least, recognize subtle secondary eclipses (see Figure \ref{cnnheatmaps}d), but it is evidently not sensitive enough to weak secondary eclipses to classify many of these simulated signals as false positives. It is possible that some of the simulated eclipsing binaries do not look like real examples from the training set (e.g. not V-shaped), or have a secondary eclipse that is too faint to detect in our representation. One way to increase our model's sensitivity to eclipsing binaries would be to include secondary and tertiary local views to give the model a ``close up'' view of any putative secondary eclipses and avoid information about the shape of secondary events from being lost in coarsely-binned global views. Including these simulated eclipsing binaries in our training set would likely help as well.

\subsubsection{Inverted Group}

We find that our model predicts about 3.2\% of the false positive TCEs in the Inverted Group to be planets with probability greater than 0.5, compared to the Robovetter, which only passes about 0.4\%. One thing we discovered when examining these false positives is that our light curve ``flattening'' technique (Section \ref{lightcurves}) occasionally makes non-transit-like phenomena look like transits. Specifically, the issue arises when we remove ``in-transit'' points before fitting the basis spline and linearly interpolate over those points. If the light curve sits below the linear interpolation then the flattened light curve gets an artificial dip that sometimes looks like a transit. We mainly observed this in light curves with high-frequency stellar variability. We originally decided to remove in-transit points to prevent the basis splines from being distorted by real transits. We could mitigate this issue in the future by improving our flattening routine, for example by interpolating non-linearly over in-transit points, or including views of the data with alternative detrending methods, which might give different or better representations of the putative transits. We also noticed that some of the model's confident predictions were spurious signals that had some features in common with real planet candidates, but were easily distinguishable by eye. Signals like these were not present in our training set, which limited our model's ability to properly classify them as false positives. Our model would likely benefit from including this set of false positives in our training set, to further improve its ability to reject spurious signals.

\subsubsection{Summary}

There are several caveats to keep in mind when using these simulated data to assess our model's performance. Firstly, inverted light curve tests do not reproduce all known types of false positives (e.g. those caused by sudden dropouts in pixel sensitivity caused by cosmic rays), so they do not assess our model's ability to identify these signals. On the other hand, inverted light curve tests may introduce new types of false positives not present in normal transit searches or our training set, which our model is not well equipped to identify. Secondly, some of the simulated false positives may not be physically realistic. For example, in many cases, eclipsing binaries with nearly equal primary and secondary eclipse depths will have V-shaped eclipses, which is not necessarily reproduced in these tests. Finally, we have not yet used these simulated data to train our model, as was done for the Robovetter (whose metrics were tuned for optimal performance on the simulated data, \citealt{koi8}). The performance of our model would likely substantially improve if we were to include these simulated data in our training set. 

%When comparing our model to the Robovetter using these simulated data, it is important to note that the Robovetter's metric thresholds were tuned to maximize the number of TCEs in Injected Group 1 classified as planet candidates, and minimize the number of TCEs from the Inverted Group classified as planet candidates \citep{koi8}.

%or example, some simulated eclipsing binaries have very large secondary eclipse ratios that may be outside the range of any real examples in the training set.
%It's actually fairly common for there to be extremely shallow secondary eclipses in light curves -- but I think you're right that the model may be recognizing that the shapes of eclipsing binaries and the ratio of the primary to secondary depth don't match what it usually sees. I have emailed Jessie to find out exactly what she did to create the injected models. 

All in all, while our model in general performs well on these tests, it does not yet match the performance of the more mature Robovetter in identifying simulated false positives. The failure modes we have observed in these tests highlight limitations in our current model implementation, which will have to be addressed before our model is ready to blindly classify TCEs to determine planet occurrence rates.  We are planning to work on improving our model's performance on these types of false positives, and we discuss some avenues for future work in Section \ref{discussion}. In the meantime, our model is still able to quite accurately sort true planets from planet candidates (even if it does require some human supervision to assure reliable classifications). We take advantage of this in Section \ref{newcandidates} when we use our model to sort TCEs from a new transit search and identify the most likely new planets. 

%stars can be determined using pixel-level data to compute the ``centroid offset'' -- that is, how the apparent centroid of the target star changes during the transits. Our model does not have access to centroid information, so it scores most TCEs in Injected Group 2 as planets with high confidence

\subsection{Comparisons to Other Automatic Vetting Systems}

\subsubsection{Robovetter}

We computed the performance of the Robovetter on our test set as an additional way of comparing it with our model. To do this, we matched the latest Robovetter classifications from the DR25 Kepler Object of Interest (KOI) table \citep[][]{koi8} with our training labels from the DR24 Autovetter Planet Candidate Catalog \citep[][]{autovettercat}. Note that the Robovetter only promotes the most plausible subset of TCEs to KOIs \citep{koi8}, so the DR25 KOI table contains only a subset of our test set (661 out of 1,523). We assumed that all missing TCEs were classified by the Robovetter as false positives. However, it is possible that some of the missing TCEs were not detected by the DR25 version of the \Kepler\ pipeline due to changes in its Transiting Planet Search module between DR24 and DR25. We tested this by computing the accuracy of the DR24 Robovetter \citep{koi7}, and found that it was nearly identical to the value we compute for the DR25 Robovetter.
% (we cannot compute AUC for the DR24 Robovetter, because it does not contain a measure of its relative confidence that each TCE is a planet)

To compute the Robovetter's metrics, we assumed our training labels were the ground truth\footnote{We make this assumption because our labels were assigned by human reviewers \citep{autovettercat}. Some TCEs in our test set appear to be misclassified, but usually our model and the Robovetter agree on these TCEs.}. We used the Robovetter's ``disposition score'' to rank the TCEs and assumed that any TCEs missing from the DR25 KOI table were rejected by the Robovetter with disposition score zero. Under these assumptions, the Robovetter has AUC of 0.974 (compared to our model's 0.988) and accuracy of 0.974 (compared to our model's 0.960).

\subsubsection{Autovetter}

The Autovetter \citep{autovetter, autovettercat} is a machine learning system for classifying \Kepler\ TCEs into 3 classes (planet candidates, astrophysical false positives and non-transiting phenomena). Most of its input features are derived from outputs of the \Kepler\ pipeline, including fitted transit parameters (e.g. period), stellar parameters (e.g. effective temperature), and signal-to-noise statistics (e.g. MES). An advantage of this approach is that features can be chosen that are known {\em a priori} to correlate with one of the output classes. For example, the Autovetter has a feature that indicates whether a TCE's period is close to the period of another TCE on the same star, which would indicate that the TCE is an astrophysical false positive caused by starspots or an eclipsing binary. A disadvantage is that the Autovetter depends on the \Kepler\ pipeline. We could not have used the Autovetter to rank our new TCEs in Section \ref{newcandidates} because we discovered those TCEs using an alternative detection pipeline.
%A similar feature tests whether a TCE's period and epoch is close to that of another TCE on any other star; this could indicate that the signal was caused by light contamination from another \Kepler\ target.

% Our model can be used to score any \Kepler\ TCE, regardless of the detection pipeline.

%The Autovetter is also likely to be tied to a specific version of the \Kepler\ pipeline. For example, if the \Kepler\ pipeline's transit-fitting model was improved, or if it changed to a more fine-grained search of period, duration and epoch (which could increase its measured MES), the Autovetter would likely need to be retrained on the new outputs.

Our approach differs from the Autovetter in that we directly model the shapes of light curves, rather than statistics derived from light curves and other sources. However, the two approaches are not mutually exclusive: the Autovetter includes the \textit{LPP} metric \citep{lpp}, which encodes information about light curve shape, and we intend to experiment with auxiliary input features in our model in the future (see Section \ref{future}).

% We would not have been able to use the Autovetter for , because we use the Box Least Squares algorithm \citep{kovacs} but the Autovetter relies on outputs from the \Kepler\ pipeline. Indeed, the Autovetter is likely to depend on the version of the \Kepler\ pipeline. [input distribution changes].

The Autovetter's accuracy on its training set is 0.973, or 0.986 if its classification objective is simplified to ``planet'' versus ``not planet'' \citep[][]{autovettercat}. This is not directly comparable to our model's test set accuracy because the Autovetter re-weights its random forest votes to minimize the total number of misclassifications across its training set \citep[][]{autovettercat, autovetter}.

Empirically, we found that the main difference between our model's classifications and the Autovetter's classifications is on the set of false positive TCEs whose signals originate on stars other than the intended target. We found that our model misclassifies many of these TCEs as planets (see Section \ref{inj2}), whereas the Autovetter correctly classifies them as false positives. The Autovetter has multiple features related to centroid shifts during transits, but our model does not currently take any information about the location of the transits on the pixels. We plan to incorporate this information into our model in the future (see Section \ref{future}).

\subsubsection{\ron \citet{armstrong}}

% The \textit{LPP} metric developed by \citet{lpp} is based on an unsupervised machine learning model that clusters light curves with similar shapes. It is used by the Robovetter and Autovetter to identify TCEs with ``not transit-like'' shapes, rather than providing a complete classification on its own. In particular, it does not identify eclipsing binary false positives. We do not compare it directly to our model.

\citet{armstrong} present a method for classifying \Kepler\ TCEs using an unsupervised machine learning algorithm to cluster light curves with similar shapes. They report accuracy of 0.863 on \Kepler\ planet candidates (compared to our model's 0.949), and 0.875 on \Kepler\ false positives (compared to our model's 0.963).

% The multiple event statistic \citep[MES][]{jenkins-mes} measures the strength of a transit signal relative to the noise.
% in a light curve.
% The multiple event statistic \citep[MES][]{jenkins-mes} measures the significance of a detected transit in a light curve. It depends on the signal-to-noise of the transits within the flux variability of the target star. The \Kepler\ pipeline has a threshold to only output TCEs with measures MES \textgreater 7.1 in order to minimize the number of false positives due to random noise \citep{TODO}.
% The strength of injected signals ranges from very strong to very faint. 

\section{Testing on New Candidates}\label{newcandidates}
\subsection{Transit Search}\label{vetting}

We tested the effectiveness of our new vetting tools by conducting a search of known \Kepler\ multi-planet systems for new TCEs and planet candidates. Known multi-transiting-planet systems are fertile grounds for detecting new transiting planets. We know {\em a priori} both that these systems have edge-on inclinations and that the planets in these systems very likely have small mutual inclinations, as is required for the known planet candidates to transit. It is likely that in many of these systems there are undiscovered planets that have not yet been detected. 

We chose to search a subset of the \Kepler\ multi-planet candidates. We searched 455 two-candidate systems, 141 three-candidate systems, 50 four-candidate systems, 21 five-candidate systems, 2 six-candidate systems, and 1 seven-candidate system. We used light curves produced by the \Kepler\ pipeline and processed by its Pre-search Data Conditioning (PDC) module. We performed the search using the transit search pipeline described by \citet{v16} for transit searches in K2 data\footnote{Because our pipeline was originally developed to search for planets in data from the  K2 mission, it is optimized for detecting the shorter ($\lesssim$ 40 days) period planets commonly found by K2, so most of our new detections were at short orbital periods.}. In brief, our pipeline flattens the light curves by fitting a basis spline to the light curve with outlier rejection and uses a Box Least Squares \citep[BLS,][]{kovacs} periodogram search to detect transiting planets. Our pipeline searches for points in the BLS periodogram with a signal-to-noise ratio greater than a pre-determined cutoff. If any part of the BLS periodogram has high enough signal-to-noise to warrant further scrutiny, the pipeline automatically tests for and rejects various types of false alarms before promoting detections to the status of TCEs. Upon detecting a TCE, the pipeline fits a transit model to the TCE, saves the best-fit parameters, and removes the region around transit from the light curve before re-calculating the BLS periodogram on the clipped light curve to attempt to detect additional TCEs around that same star. 

We removed transits of previously known planets from the \Kepler\ multi-planet system light curves and searched for new candidate signals. For this search, we set a signal-to-noise (S/N$_{\rm BLS}$) threshold for detecting and recording TCEs of S/N$_{\rm BLS}$ = 5. This S/N$_{\rm BLS}$ threshold is considerably more liberal than both thresholds typically used for transit searches in data from the K2 mission (which we normally set at S/N$_{\rm BLS}$ = 9, \citealt{v16}), and the threshold\footnote{Our calculation of S/N$_{\rm BLS}$ is roughly equivalent to the MES calculated by the \Kepler\ pipeline.} used by the \Kepler\ pipeline of MES = 7.1 \citep{jenkins-mes}. Traditionally, higher S/N$_{\rm BLS}$ thresholds are used for two reasons: a) to prevent the number of spurious TCEs (caused by systematics or otherwise) from overwhelming the number of real detected planets, and b) to ensure that no TCEs produced by purely Gaussian fluctuations are detected \citep[e.g.][]{jenkins-mes}. The main benefit to using a cutoff as low as S/N$_{\rm BLS}$ = 5 is that we greatly increase our sensitivity to TCEs with S/N$_{\rm BLS}$ {\em above} more conservative cutoffs, because a variety of factors can cause S/N$_{\rm BLS}$ to be underestimated and TCEs to be erroneously rejected (see Section \ref{future}). Therefore, our search had the potential to discover new planets with S/N$_{\rm BLS}$ slightly above the traditional \Kepler\ cutoff of 7.1. 

The trade-off of choosing such a low S/N$_{\rm BLS}$ cutoff is that we greatly increase the number of spurious TCEs returned by our transit search. We added post-processing logic to automatically remove two common sources of spurious detections:
\begin{enumerate}
  \item We discarded any detected TCEs with transit duration less than 29.4 minutes (the spacing between consecutive \Kepler\ data points). With a low S/N$_{\rm BLS}$ threshold, our pipeline tends to find a large number of ``transits'' consisting of only a couple of points, which would translate to extremely short transit durations that are not physically realistic.
  \item We discarded any TCEs with fewer than 3 complete transits. Often, one or two spurious events (like instrumental noise or cosmic ray hits) lined up with one or two gaps in the light curve, causing the BLS algorithm to detect a long-period TCE. We required TCEs to have at least 3 transits containing at least half the number of expected in-transit points. Even if some of these discarded TCEs were caused by real planets, we would not have been able to confidently validate them with fewer than 3 complete transits.
\end{enumerate}

After discarding these types of false positives, our search of 670 stars returned 513 new TCEs\footnote{On occasion, our pipeline detected residuals from planet candidates we attempted to remove, producing a TCE with the same orbital period and time of transit as the previously known candidates. We identified these cases by matching transit times and periods with known candidates and ignore them in the following paragraphs.} -- nearly one TCE per star. While the number of TCEs produced by this search is still manageable because we only searched a relatively small number of stars, such a high TCE rate would make it impractical for manual vetting of a similar search of the entire \Kepler\ dataset. Therefore, we used our neural network model to sort the TCEs by the probability of being planets. Out of the 513 TCEs produced by our search, our model only predicted 30 were planets with a score greater than 0.5. These 30 highly ranked TCE are shown in Table \ref{classtable}. We also inspected each TCE by eye to confirm that the ranking of TCEs by our model appeared reasonable. Empirically, our model did a good job of ranking the TCEs, with very few obvious errors. 

\subsection{Summary of New Highly-Ranked TCEs}

Of the 30 TCEs our model classifies as more-likely-than-not planets, we give more serious consideration to the 9 TCEs with planet predictions above 0.8. On our test set, the signals our model classified as planets with probability greater than 0.8 turned out to be planets or candidates 94\% of the time. We calculated the signal-to-noise ratio (S/N) of these TCEs by measuring the mean depth of transit, dividing by the scatter of each individual point, and multiplying by the square root of the number of points in transit. All of the TCEs with scores above 0.8 have S/N above the traditional \Kepler\ cutoff of 7.1. {\ron We summarize these these TCEs with predictions greater than 0.8 in Table \ref{table:topcand} and give a brief description of each below.}

\begin{deluxetable}{cccccc}[t!]
\tablewidth{250pt}
\tablecaption{\ron Summary of New Highly-Ranked TCEs \label{table:topcand}}
\tablehead{
    \colhead{KIC ID$^\star$} &
    \colhead{KOI ID$^\dagger$} &
    \colhead{Kepler Name} &
    \colhead{Period} &
    \colhead{S/N} &
    \colhead{Prediction} \\
    & & & \colhead{\it days} & &
}
\startdata
11442793 & 351  & \Kepler-90   & 14.4 & 8.7  & 0.942 \\
8480285  & 691  & \Kepler-647  & 10.9 & 8.4  & 0.941 \\
11568987 & 354  & \Kepler-534  & 27.1 & 9.8  & 0.920 \\
4852528  & 500  & \Kepler-80   & 14.6 & 8.6  & 0.916 \\
5972334  & 191  & \Kepler-487  & 6.02 & 10.2 & 0.896 \\
10337517 & 1165 & \Kepler-783  & 11.1 & 9.6  & 0.860 \\
11030475 & 2248 &              & 4.75 & 8.7  & 0.858 \\
4548011  & 4288 & \Kepler-1581 & 14.0 & 9.7  & 0.852 \\
3832474  & 806  & \Kepler-30   & 29.5 & 14.1 & 0.847
\enddata
\tablenotetext{}{Summary of new \Kepler\ TCEs predicted to be planets by our model with probability greater than 0.8. See Table \ref{classtable} for full details.}
\tablenotetext{$\star$}{\Kepler\ Input Catalog identification number.}
\tablenotetext{$\dagger$}{\Kepler\ Object of Interest identification number.}
\end{deluxetable}

\begin{itemize}
    \item {\ron \textbf{KIC 11442793}}. This system hosts {\em seven} transiting planets, ranging from slightly larger than Earth to Jupiter-sized planets, all of which have been either validated or confirmed via transit-timing variations \citep{cabrerak90, schmitt, lissauermultis2}. We detect an additional TCE with a period of 14.4~days with S/N of 8.7. 
    \item {\ron \textbf{KIC 8480285}}. This system hosts two planets - one previously validated Earth-sized planet \citep{morton16} and one candidate super-Earth in periods of 16.2~days and 29.7~days respectively. We detect an additional TCE at S/N = 8.4 with a period of 10.9~days. 
    \item {\ron \textbf{KIC 11568987}}. This system hosts one validated super-Earth \citep{morton16} and one candidate Earth-sized planet in 16 and 7.3~day periods, respectively. We detect a new TCE with a period of 27.1~days, S/N = 9.8 and a depth corresponding to a roughly Earth-sized planet. 
    \item {\ron \textbf{KIC 4852528}}. This system hosts five confirmed small planets, the outer four of which are in a rare dynamical configuration with the middle three planets and the outer three planets each locked in a three-body resonance \citep{lissauermultis2, macdonald}. We find a new TCE with a period of 14.64~days at S/N = 8.6 exterior to all five previously known planets.  
    \item {\ron \textbf{KIC 5972334}}. This system has two validated planets \citep{morton16} and two candidates, including a warm Jupiter, an ultra-short-period planet \citep{sanchisojeda2}, and a super-Earth in between, giving it a distinct architecture reminiscent of the 55 Cnc \citep{nelson} and WASP-47 \citep{becker} planetary systems. We detect a strong (S/N = 10.2) TCE with a period of 6.02~days which, if real, would have a radius a bit larger than that of the Earth. We see extra scatter in the light curve during the transit of this new TCE, however, which needs to be better understood before this TCE can be considered a good candidate.
    \item {\ron \textbf{KIC 10337517}}. This system hosts a sub-Earth-sized validated planet in a 4.3~day orbit \citep{morton16} and a super-Earth-sized planet candidate in a 7~day orbit. We detect an additional TCE in an 11.1~day orbit, which also would likely be sub-Earth-sized. 
    \item {\ron \textbf{KIC 11030475}}. This system has 4 planet candidates, none of which have been validated. We detect a new TCE at an orbital period of 4.75~days and S/N =  8.7, with the same time of transit, depth, duration, and exactly half the period of a previously detected candidate, KOI 2248.02. This ``new'' TCE, therefore, appears to be the true period of the candidate KOI 2248.02.
    \item {\ron \textbf{KIC 4548011}}. This system hosts one validated planet \citep{morton16} in a 6.3~day period and one additional candidate in a 9.1~day period. We detect a new TCE with a period of 13.95~days with a relatively strong S/N = 9.7, which if real, would likely be a sub-Earth-sized planet. Subsequently, this signal was also detected by the \Kepler\ pipeline and included in the DR25 release of the \Kepler\ planet candidate catalog as KOI 4288.04.
    \item {\ron \textbf{KIC 3832474}}. This system hosts two confirmed giant planets, and a smaller confirmed super-Earth-sized planet called \Kepler-30 b, which orbits with a period of 29.16~days \citep{fabrycky}. All of these planets show transit timing variations (TTVs), and \Kepler-30 d shows extremely large TTVs with an amplitude of more than a day \citep{fabrycky,kepler30}. We detect a new TCE with a period of 29.45~days - very close to the mean period of \Kepler-30 d. We suspect this TCE is due to residual transits of \Kepler-30 d that we were unable to completely remove from the light curve due to the large TTVs. 
\end{itemize}

Four of these new TCEs were {\ron predicted to be planets by our model with probability 0.9 or greater.} We consider this to be a very confident prediction -- among those objects in our test set to which our model assigned a planet probability greater than 0.9, 96\% turned out to be correctly classified as planets. {\ron Given our model's high confidence in these four new candidates,} and the fact that they all have signal-to-noise ratios greater than the traditional \Kepler\ threshold of S/N = 7.1, we subject these four new TCEs to additional scrutiny in order to validate them as {\em bona fide} exoplanets. We describe our additional tests and calculations to this end in the following subsections. 

\subsection{Vetting New TCEs}

\subsubsection{Nearby Stars to the New TCE Hosts}

First, we attempted to identify whether any stars have been detected close on the sky to the host stars of our four new TCEs with planet predictions greater than 0.9. It is important to identify nearby stars and rule out the possibility that these companion stars are the sources of the new transit signals we see. To do this, we used information uploaded to the \Kepler\ Community Follow-up Program (CFOP) webpages\footnote{\url{https://exofop.ipac.caltech.edu/cfop.php}} for these four targets. 

CFOP reports that there is a companion star about seven arcseconds south of \Kepler-90, detected in imaging from the United Kingdom Infra-Red Telescope (UKIRT). This star is bright enough to potentially be the source of the transit, and its nearby position overlaps \Kepler-90's point spread function, so it could in principle be the source of the new shallow transit signal we see. Similarly, UKIRT revealed a faint companion about 7~arcseconds southeast of \Kepler-647 which could contribute the transit signal. Near \Kepler-534, there are six stars within three \Kepler\ pixels that could in principle contribute the transit signals we see, and Howard Isaacson reports on CFOP the detection of a faint secondary set of spectral lines in a high resolution spectrum taken at Keck Observatory \citep[these lines were identified using the method of][]{reamatch}. Finally, \citet{kraus} report that \Kepler-80 hosts two faint companions to the South - one at a distance of 1.7~arcseconds, and one at a distance of 3.9~arcseconds, both of which could in principle be the sources of the new transit signal we see.  

Most of these detected companions, in particular, the ones detected visually in UKIRT or adaptive optics imaging, are at projected distances far enough from the \Kepler\ target stars that we can either show that they are the sources of the transits we see, or rule them out as the sources of the transits. The spectroscopically detected companion to \Kepler-534, however, is apparently close enough to the target star that it a) contributed light to the Keck HIRES spectrograph's slit, and b) was not detected with Robo-AO and Palomar Observatory adaptive optics imaging \citep{ziegler, jiwang, furlan}. Since we do not know this secondary star's angular separation, we therefore cannot rule it out as the source of the new transiting signal we see, and we do not proceed to statistically validate this TCE.

\subsubsection{Image Centroid Analysis}

Each of the three remaining high-quality TCE hosts have nearby visually detected companions within their target apertures that might cause the transits. In this subsection we use the lack of apparent motion of the positions of the target stars in the \Kepler\ images to show that the transits must be co-spatial with the target stars and cannot be contamination from these nearby visually detected companions.

We started by downloading the long-cadence target pixel time-series files for these three objects. For each of the roughly 60,000 images for each of the three stars, we measured the position of the target star by fitting a two-dimensional Gaussian to the image, and recorded the best-fit center position in both the X and Y dimensions. This yielded a time series of X and Y positions for the entire four years of the \Kepler\ mission. About every 90 days, \Kepler\ rotated to keep its solar panels pointed towards the sun, which meant that the stars fell on different cameras and the pixels were in different orientations every 90 days\footnote{These 90 day periods are called ``quarters''.}. We therefore transformed the X and Y pixel time series into Right Ascension (R.A.) and Declination (Dec.) time series. We high-pass-filtered the R.A. and Dec. time series for each star, identified the points that occurred during the transits of the three TCE signals, and calculated the average change in R.A. and Dec. for the target star during transit of the new TCEs. 

For all three of our targets, we found no significant motion of the centroid during transit. For \Kepler-80, we calculated a mean shift during transit in R.A. of 0.1~milliarcseconds and a shift in Dec. of 0.05~milliarcseconds, each with uncertainties of 0.13~milliarcseconds. If the transits, which have a depth of about 250 parts per million, were in fact coming from the closest visually detected companion at a distance of 1.7~arcseconds, the expected centroid shift would be 1700~milliarcseconds $\times 2.5\times 10^{-4} = 0.42$~milliarcseconds, smaller than what we measure. We conclude the new TCE around \Kepler-80 is co-spatial with the target star (with an uncertainty of about 0.6~arcseconds). Similarly, we find no shift in the centroid position of \Kepler-90 during the transit of the new TCE with uncertainties of about 0.1~milliarcsecond. Therefore, a roughly 90 ppm transit like our new TCE must come from a co-spatial source with uncertainties of about 1~arcsecond, which excludes the closest detected companion star (7~arcseconds distant) as the potential source of the transits. Finally, we confirm that the source of the new TCE around \Kepler-647 must be co-spatial with the target star with an uncertainty of roughly 1.5~arcseconds - much closer than the closest detected companion at a distance of 7~arcseconds. 

We therefore conclude that none of the visually identified companions to the three stars hosting new high-quality TCEs are the sources of the new transit signals we see.

\subsubsection{Photometric Tests}

In this subsection, we describe the tests and checks we performed on the \Kepler\ light curves of the TCE host stars to make sure that the TCE signals are not due to instrumental artifacts or other false positive scenarios. In particular, we performed the following checks and tests: 

\begin{enumerate}
    \item {\ron \textbf{Secondary eclipse search.}} We inspected the phase-folded \Kepler\ light curves for hints of a secondary eclipse. Detecting a secondary event would indicate that the TCE is likely due to the decrease in light caused by a faint, co-spatial star that happens to be an eclipsing binary. We see no evidence for a significant secondary eclipse for any of our targets. 
    \item {\ron \textbf{Different apertures.}} We produced light curves from the pixels outside of the optimal \Kepler\ aperture for each of our targets and inspected these light curves folded on the period of the TCE signals. If the transits are present in the background pixels with a depth considerably greater than we see in the on-target pixels, the signals are likely being introduced to the targets through scattered background light. In all three cases, we found that the background pixels showed no significant signals, so the TCEs are consistent with having their source on or near the target star. 
    \item {\ron \textbf{Even-odd transits.}} We checked to make sure that the depths, timings, and shapes of even and odd-numbered transits were consistent. Detecting a significant difference between even and odd numbered transits would indicate that the signal is caused by a fainter eclipsing binary star with twice the period that we detect, and with slightly different primary and secondary eclipses. We see no evidence for even/odd differences in the three targets we inspected. 
    \item {\ron \textbf{Consistency between quarters.}} We checked that the transit signals did not change significantly depending on which CCD module the target stars were observed. Every quarter during its original mission, \Kepler\ rolled to keep its solar panels facing the sun, so every roughly 90 days, the stars fell on different positions on the CCD array. Sometimes, contaminating signals could be introduced into the light curve more strongly in some positions on the CCD than others, causing differences in the strengths and depths of transit signals on each CCD.  We saw no evidence for these quarterly transit depth differences. 
    \item {\ron \textbf{Consistency over mission lifetime.}} We checked that the transit signals did not change significantly between the first and second halves of the \Kepler\ mission. Sometimes, contamination can be caused by a process called ``charge transfer inefficiency'', or ``column anomaly'' \citep{coughlinephemeris}. In this scenario, a variable star (likely an eclipsing binary) that happens to be on the same CCD column as the target star can bleed signal during readout of the images. This effect often worsens over time \citep{koi7}, so often, the transit depths at the beginning of the mission are considerably shallower than the transit depths at the end of the mission \citep[see, for example, Figures 5 and 6 of][]{gaidoscolumn}. We see no significant differences in transit depth in the light curves of the new TCEs around \Kepler-80 and \Kepler-90, making it unlikely these two TCE signals are caused by column anomaly.  However, we do see a possible increase in transit depth for the new TCE around \Kepler-647. We fit the increase in transit depth as a linear function of time, and determined a slope of about 13 $\pm$ 5 ppm per year -- inconsistent with a constant transit depth with about $2.6\sigma$ confidence. While the detected slope is weak, and could plausibly be due to red noise in the light curve or underestimated uncertainties in the transit depths, out of an abundance of caution, we choose not to validate this TCE.
    \item {\ron \textbf{Ephemeris matching.}} We searched for ``ephemeris matches,'' or other targets with nearly identical periods and times of transits either observed by \Kepler\ or falling near the \Kepler\ field in the sky. Finding an ephemeris match is an indication that a signal is likely caused by contamination from either the matched source, or from another unknown astrophysical source. We searched through the list of periods and transit times collected by \citet{coughlinephemeris}. We found no plausible other sources for the TCE signals in question (including the TCE around \Kepler-647, which we suspect might be due to column anomaly). 
    \item {\ron \textbf{Simple Aperture Photometry.}} We confirmed that the transit signals were still present in the raw Simple Aperture Photometry (SAP) time series, and that their characteristics did not substantially change between the SAP photometry and the photometry corrected for systematics via PDCMAP. This shows that the transit signals are likely not residuals of larger systematic errors in the \Kepler\ light curves that were partially corrected by PDCMAP.  
\end{enumerate}

In the end, we find that the two new TCE signals around \Kepler-80 and \Kepler-90 appear to be astrophysical and not caused by \Kepler\ systematics, and we find no evidence that these TCEs are associated with other known astrophysical signals from elsewhere in the \Kepler\ field of view. We therefore consider these two signals to be planet candidates. While the TCE around \Kepler-647 passes most tests, we find that the transit depth appears to increase over the course of the mission -- a possible sign that this signal is an astrophysical false positive caused by column anomaly. While this candidate could still be a planet, we are not confident enough in this interpretation to validate it with high confidence. 

\subsection{False Positive Probability Calculations}

After we vetted the two candidate signals around \Kepler-80 and \Kepler-90 to rule out various instrumental and astrophysical false positive scenarios, we calculated the probability that these two signals are caused by other astrophysical false positive scenarios that are indistinguishable from the signals we see, given the data at our disposal. We performed this calculation using \vespa\  \citep{vespa}, an open source code library that implements the methodology described by \citet{morton12} and \citet{morton16}. In brief, \vespa\ considers the possibility that the signal is caused by a transiting planet, as well as several different astrophysical false positive scenarios, including the possibility that the signal is caused by a binary star eclipsing the target, or that the signal is caused by some binary star around another, fainter star nearby on the sky to the target. After measuring basic transit parameters and fitting the host star's spectral energy distribution, \vespa\ calculates how well these parameters match the various scenarios it considers, and calculates the relative likelihood of each scenario using priors based on observational constraints and the characteristics of the host star. 

We ran \vespa\ on the new candidates around \Kepler-80 and \Kepler-90. We included as constraints the contrast curves from speckle imaging observations uploaded to CFOP by Mark Everett (two observations taken at 692 and 880~nm at Gemini-N Observatory of \Kepler-90 and one observation taken at 880~nm at the WIYN telescope of \Kepler-80). We imposed constraints on the depths of putative secondary eclipses by calculating the $3\sigma$ limit based on the scatter in the phase-folded light curve on each candidate's period binned to the transit duration. We used our constraints from the image centroid analysis to specify the $3\sigma$ maximum radius at which a contaminating star could contribute the transit signal. We used the spectroscopic parameters from \citet{macdonald} and \citet{petiguracks} as inputs for \Kepler-80 and \Kepler-90 respectively, and we included near infrared photometry from the 2MASS survey \citep{Skrutskie:2006}. Recently, it has been shown that systematic photometric uncertainties and/or offsets between photometric datasets between can cause \vespa\ to mis-classify planet candidates (\citealt{shporer2017}, Mayo et al., in prep), so we avoided this issue by only including the high-quality 2MASS photometry. 

Based on these inputs, \vespa\ calculated that the new candidate around \Kepler-80 has a false positive probability of about 0.002, and the new candidate around \Kepler-90 has a false positive probability of about 0.005. Both of these false positive probabilities are quite low, below the 1\% threshold often used to validate planets in \Kepler\ \citep[e.g.][]{morton16}, thanks to the well characterized stars, high-resolution imaging, and tight constraints on the source of the transits from our centroid analysis. The \vespa\ anaysis assumes that the candidates being validated are the only candidates in their systems, but these two candidates are found multi-planet systems. Candidates in multi-planet systems are {\em a priori} significantly less likely to be false positives than candidates in single-planet systems \citep{lathammultis, lissauermultis1, rowe}. \citet{lissauermultis1} estimate that candidates in systems with at least three planets are about 50 times more likely to be real planets than the average planet. When we factor in this multiplicity ``boost'', the false positive probabilities of our two candidates decrease to less than 10$^{-4}$. We therefore consider these two new candidates to be validated as {\em bona fide} exoplanets, and designate them \Kepler-80~g and \Kepler-90~i.

\subsection{Transit Fits}
\begin{figure*}[t!]
\centering
\includegraphics[width=350pt]{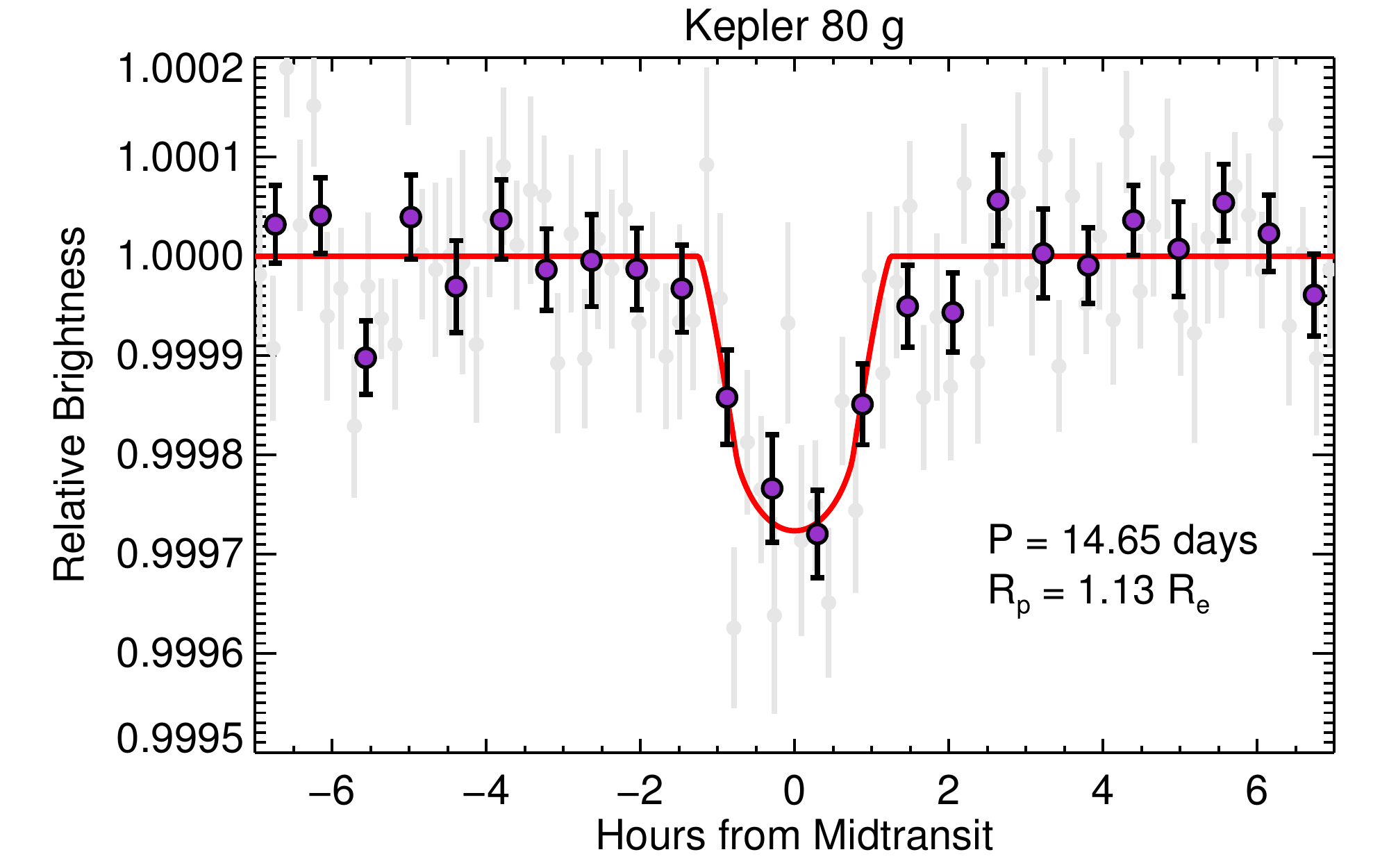}
\includegraphics[width=350pt]{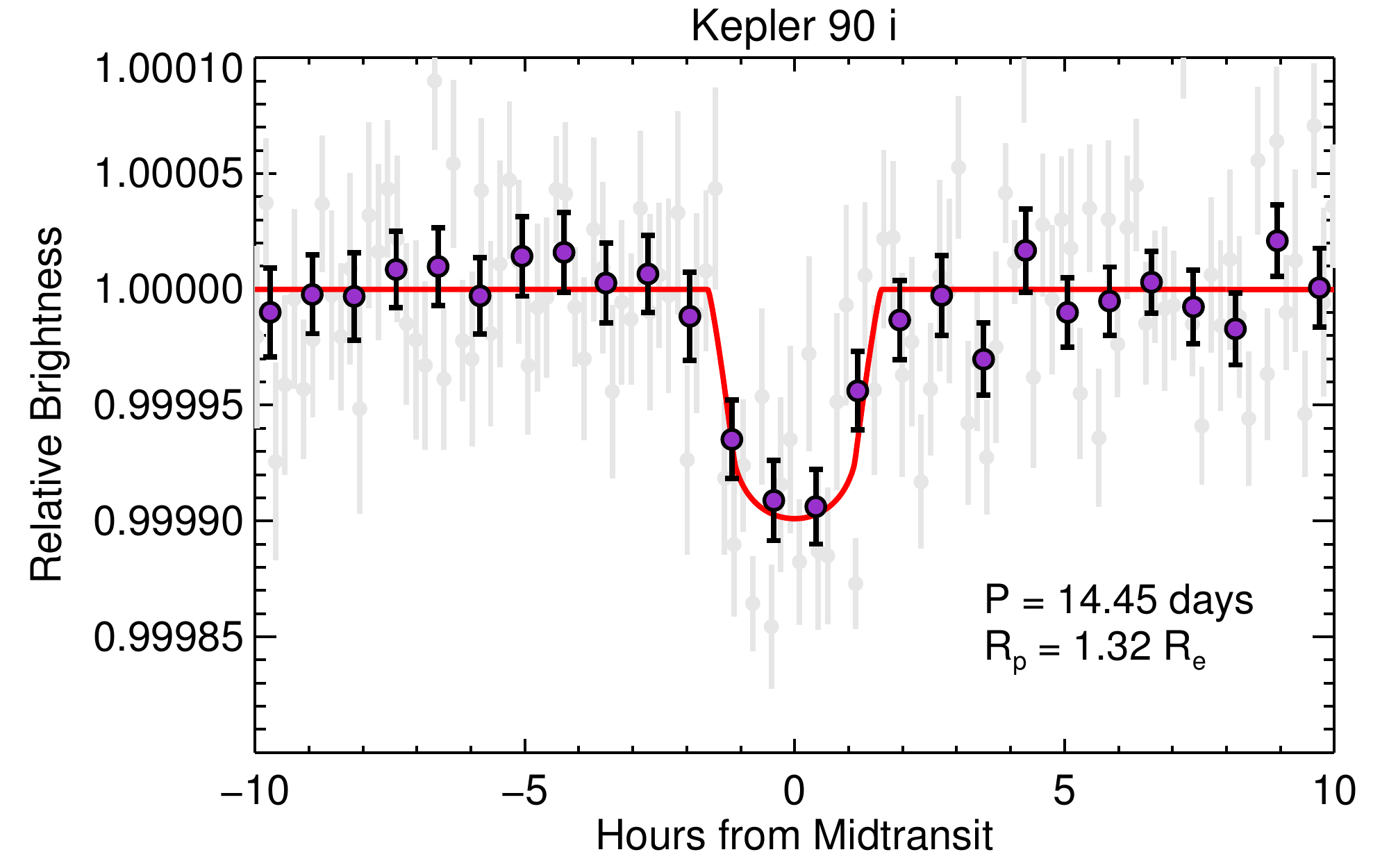}

\caption{Transit light curves and best-fit models for the {\ron newly discovered planets} around \Kepler-80 and \Kepler-90. In these plots, the grey points are robust averages of bins with width of approximately 10 minutes. The purple points are robust averages of bins with size about 1/4 the transit duration of the planet (bins of about 30 minutes for \Kepler-80~g and about 45 minutes for \Kepler-90~i.} \label{transitfigures}
\end{figure*}

After statistically validating the two {\ron newly discovered planets} around \Kepler-80 and \Kepler-90, we fit the full \Kepler\ long-cadence light curves with \citet{mandelagol} models to determine planet and orbital parameters and uncertainties. We explored parameter space using a Markov Chain Monte Carlo (MCMC) algorithm with affine invariant ensemble sampling \citep{goodmanweare}. We fit the new candidates' orbital periods, times of transit, orbital inclinations, scaled semimajor axes ($a/R_\star$), radius ratios ($R_p/R_\star$), and quadratic limb darkening coefficients \citep[following the parameterization of][]{kippingld}. We imposed Gaussian priors on the two limb-darkening coefficients, centered at values predicted by \citet{claretbloemen} for the host stars (using spectroscopic parameters from \citealt{macdonald} for \Kepler-80, and parameters from \citealt{petiguracks} for \Kepler-90). We set the widths of these Gaussian priors to be 0.07 in both the linear and quadratic coefficients, following \citet{wolter}. 

For each planet fit, we initialized 100 walkers, and evolved them for 20,000 links, discarding the first 10,000 for burn-in. We tested convergence of the MCMC chains by calculating the Gelman-Rubin potential reduction factor \citep{gelmanrubin}, which we found to be below 1.2 for all of our parameters \citep{brooks1998}. We show the results of our transit fits in Table \ref{bigtable}, and we plot the transit light curves of these two {\ron newly discovered} planets in Figure \ref{transitfigures}. 

\begin{deluxetable*}{lcc}
\tablecaption{Parameters for \Kepler-80 ${\rm g}$ and \Kepler-90 ${\rm i}$ \label{bigtable}}
\tablewidth{0pt}
\tablehead{
  \colhead{Parameter} & \colhead{\Kepler-80~g}  & \colhead{\Kepler-90~i}}
\startdata
\emph{Stellar Parameters} & \\
Right Ascension & 19:44:27.02 & 18:57:44.04  \\
Declination & 39:58:43.6 & 49:18:18.58 \\

$M_\star$~[$M_\odot$] & 0.83 $\pm$ 0.03  & 1.12  $\pm$ 0.046\\
$R_\star$~[$R_\odot$] & 0.678  $\pm$ 0.023 & 1.29  $\pm$ 0.167\\
Limb darkening $q_1$~ &  \ldoneg $\pm$ \uldoneg & \ldonei $\pm$ \uldonei \\
Limb darkening $q_2$~ &  \ldtwog $\pm$ \uldtwog & \ldtwoi $\pm$ \uldtwoi \\

$\log g_\star$~[cgs] & 4.639 $\pm$ .012 & 4.27 $\pm$ 0.1 \\
\meh & 0.04 $\pm$ 0.08 & 0.098 $\pm$0.04\\
$T_{\rm eff}$ [K] & 4540 $\pm$ 100& 6015 $\pm$ 60 \\
 & & \\
 
\emph{New Planet Parameters} & \\
Orbital Period, $P$~[days] &  \perplg $\pm$ \uperplg & \perpli $\pm$ \uperpli \\
Radius Ratio, $(R_P/R_\star)$ &   \rprstg $\pm$ \urprstg & \rprsti $\pm$ \urprsti \\
Scaled semi-major axis, $a/R_\star$  &   \arstg $\uarstg$ & \arsti $\uarsti$ \\
Orbital inclination, $i$~[deg] &   \inclg $\uinclg$ & \incli $\uincli$ \\
Transit impact parameter, $b$ &   \impg $\pm$ \uimpg & \impi $\pm$ \uimpi \\
Time of Transit $t_{t}$~[$\rm BJD_{\rm TDB}$] &   \ttransitg $\pm$ \uttransitg & \ttransiti $\pm$ \uttransiti\\ 
Transit Duration $t_{14}$~[hours] &   \tdurg $\pm$ \utdurg & \tduri $\pm$ \utduri \\ 
Equilibrium Temperature $T_{eq}$~ [K] & \teqg $\pm$ \uteqg &\teqi $\pm$ \uteqi \\
$R_P$~[\rearth] &  \rplg $\pm$ \urplg   & \rpli $\pm$ \urpli \\

\enddata

\tablecomments{Spectroscopic and stellar parameters for \Kepler-80 come from \citet{macdonald}, while the parameters for \Kepler-90 come from \citet{petiguracks} and \citet{johnsoncks}. Equilibrium temperatures are calculated assuming perfect heat redistribution and albedos uniformly distributed between 0 and 0.7. }

\end{deluxetable*}

\section{Discussion}\label{discussion}
\subsection{\ron Newly Validated Planets}

We identified two new exoplanets by conducting a transit search and using our new machine learning planet candidate vetting pipeline. Both of these planets are in high-multiplicity systems (there are five and seven other planets in the \Kepler-80 and \Kepler-90 systems, respectively), which are interesting to study because of our fine understanding of their architectures. In this subsection, we comment on each of these two planets individually. 

\subsubsection{\Kepler-80~g}

\Kepler-80~g is the outermost planet in its system. Previously, five planets were known, with orbital periods of 0.98, 3.07, 4.65, 7.05, and 9.52~days. The outer four\footnote{The shortest-period planet in the system, \Kepler-80 f, is dynamically decoupled from its resonant counterparts.} of these planets are in a rare dynamical configuration called a resonant chain \citep{lissauermultis2, macdonald}. The orbital periods of each pair of adjacent planets are close to an integer ratio (in this system, either 2:3 or 3:4), and the orbital periods of consecutive trios of planets satisfy the Laplace relationship: 

\begin{equation} \label{laplace}
    \frac{p}{P_1} - \frac{p + q}{P_2} + \frac{q}{P_3} \approx 0
\end{equation}

\noindent where $p$ and $q$ are integers, and $P_i$ are the orbital periods of the three planets. \Kepler-80 d,e, and b satisfy this relationship with $(p,q)$ = $(2,3)$, and \Kepler-80 e,b, and c satisfy this relationship with $(p,q)$ = $(1,2)$. Satisfying the Laplace relationship is equivalent to the fact the TTV ``super-period'' for adjacent planets in the chain is nearly the same for each pair of planets. In the \Kepler-80 system, the super-period of each adjacent pair of planets is close to 192 days \citep{macdonald}. The super-period is the time it takes for the orbital angle at which two adjacent planets line up in conjunction circulates about the orbit -- that is, for planets in the \Kepler-80 system, two adjacent planets which happen to transit simultaneously will repeat this every 192 days. The super-period, $P_{\rm super}$, for two planets close to a first order mean motion resonance is given by: 

\begin{equation}
    P_{\rm super} = \frac{1}{|\frac{j + 1} {P_2} - \frac{j}{P_1}|}
\end{equation}

\noindent where $j$ is an integer and $P_i$ are the orbital periods of the two planets. Resonant chains like the one around \Kepler-80 are extremely finely tuned, specifying orbital periods to high precision. Recently, this fact was used by \citet{luger} to predict and recover the orbital period of the planet TRAPPIST-1 h using data from the K2 mission. Following \citet{luger}, we calculated the orbital periods of putative additional outer planets in the \Kepler-80 system under the assumption that they happened to be in a three-body resonance with \Kepler-80 b and c. After rearranging Equation \ref{laplace} and plugging in values of $p$ and $q$ between 1 and 3 (inclusive), we identified possible orbital periods of 14.64671, 11.54078, 10.77887, 31.72031, 12.41860, and 20.04003~days. For $p = q = 1$, the period predicted by the Laplace relation is within 100~seconds of the measured orbital period (\perplg $\pm$ \uperplg~days) of the {\ron newly discovered} planet -- agreement to within one part in 10,000. \Kepler-80~g orbits just outside a 2:3 period ratio with the 9.52~day planet, and these two planets have a super-period of 192~days, matching the super-periods of other planet pairs in the system. \Kepler-80~g therefore almost certainly is part of the chain of three-body resonances around \Kepler-80, bringing the total number of planets in the chain to 5, and the total number of planets in the system to six. The fact that we can so accurately predict the orbital period of \Kepler-80~g through resonance arguments is {\ron strong} independent confirmation that the signal is indeed caused by an exoplanet, and a validation of our methods from end-to-end. 

\Kepler-80~g warrants further study for several reasons. First, although the orbital period we found makes it extremely likely this {\ron newly discovered} planet is the outer link in a chain of three-body resonances, the system requires further dynamical analysis to confirm the resonant state. Even though its low signal-to-noise ratio will make it difficult, it could be worthwhile to search for \Kepler-80~g's transit timing variations, perhaps by measuring the timing of many transits averaged together \citep[e.g.][]{mills}. Finally, it will be important to assess the impact of a new planet in the resonant chain on the \Kepler-80 TTV mass measurements made by \citet{macdonald}. In principle, if some of the TTV signal measured in the 9.52~day \Kepler-80 c was caused by \Kepler-80 g, that would impact the measured mass of \Kepler-80 b, and so on.

\subsubsection{\Kepler-90~i}

\Kepler-90 is a star that was previously known to host seven transiting planets. Planetary systems with this many known planets are very rare. According to the NASA Exoplanet Archive (accessed 19 August 2017), only \Kepler-90 and TRAPPIST-1 \citep{gillon} are known to host this many planets\footnote{HD 10180 \citep{lovis}, shows six confidently detected radial velocity signals and up to three more speculative signals \citep{lovis, tuomi} which have not been found necessary in subsequent analyses \citep{kane}.}. \Kepler-90 was initially detected by the \Kepler\ pipeline, and designated KOI~351 by \citet{koi2}, who detected three large, long-period transiting planet candidates, which eventually would be known as \Kepler-90 e, g, and h. As more data became available, additional searches \citep{cabrerak90, koi4, schmitt, lissauermultis2} identified additional shorter-period and smaller transiting planet candidates, bringing the total number of planets to seven. \Kepler-90 hosts two planets about 1.5 times larger than Earth in 7 and 8.7~day orbits, three planets about three times the size of Earth in 59.7, 91.9, and 124.9~day orbits, and two giant planets in 210.6 and 331.6~day orbits.

The {\ron newly discovered} planet, \Kepler-90~i, falls in the largest gap in orbital-period space in the \Kepler-90 system between \Kepler-90 c (8.7~days)  and \Kepler-90 d (59.7~days). With a period of 14.4~days and a radius of 1.3~\rearth, \Kepler-90~i fits in well with \Kepler-90 b and c, the two shortest period planets in the system, which have radii comparable to \Kepler-90~i of 1.4 and 1.6~\rearth\, respectively. \Kepler-90 b, c, and i all have radii small enough that they are either likely or might be rocky \citep{rogers}. It is striking that the planets around \Kepler-90 are so well sorted by radius and orbital period, with small planets orbiting close-in and giant planets orbiting farther from the star, although it is possible that this is a detection bias. \Kepler-90~i is just barely above \Kepler's detection limits, so we would be unlikely to detect any similarly small planets orbiting farther from \Kepler-90 than this.  

Another particularly striking feature of the \Kepler-90 system is how orderly it appears to be. The transit durations $t_{1,4}$ of the seven previously known \Kepler-90 planets scale quite well with period $P$ (where $t_{1,4} \propto P^{1/3}$). \Kepler-90~i appears to somewhat buck that trend. {\ron While the mean stellar density derived from \Kepler-90~i's transit is still consistent with the other planets and \Kepler-90's spectroscopic density,} \Kepler-90~i's transit duration of about 2.8~hours is shorter than we expect (about 5~hours) assuming circular and co-planar orbits. Therefore, \Kepler-90~i is likely slightly inclined away from our line of sight. If the orbit of \Kepler-90~i is perfectly circular, its impact parameter would have to be about 0.85, giving it an orbital inclination of about 88~degrees, compared to the other seven planets in the system, which all likely have orbital inclinations 89~degrees or higher. There are reasons to expect that the \Kepler-90 system should host some planets with slightly higher inclinations than are determined for the seven previously known planets -- \citet{beckeradams} showed that the \Kepler-90 system was among the \Kepler\ multi-planet systems where self-excitation of mutual inclinations was most likely to occur. The addition of an eighth planet would make it even more likely that planets might be excited to impact parameters of $\approx$ 0.85 as we see for \Kepler-90 i. 

The addition of an eighth planet makes \Kepler-90 a record-setter as the first exoplanetary system known to have eight planets, and along with the Sun, only one of two stars known to host this many planets. Finding systems with this many planets and learning about their sizes and orbits is a unique window into their formation. As noted by \citet{cabrerak90} and \citet{schmitt}, \Kepler-90 has an architecture reminiscent of the Solar system, with smaller planets orbiting close to the host star, and larger planets orbiting at larger distance. Unlike the Solar system, however, the entire \Kepler-90 system is packed within about the orbital radius of Earth. This compact architecture suggests the \Kepler-90 planets may have formed like the Solar system planets, with the more distant planets forming beyond the snow-line, where heavy volatile elements like water are in solid phase, and with the smaller planets forming closer to the star, where heavy volatile elements are in gaseous phase, limiting their growth. Then, unlike the solar system, \Kepler-90 may have undergone disk migration to bring it into its current flat, compact, and orderly configuration. If this is true, then there might be more un-detected transiting planets orbiting \Kepler-90, which have radii too small or orbital periods too long to detect without more \Kepler-like observations.  

\subsection{Looking Ahead}\label{future}

The goal of the \Kepler\ mission was to study Earth-sized planets orbiting Sun-like stars. However, even with 4 years of continuous data, these planets are difficult to detect because: a) their smaller radii and the relative dimness of their stars means the dips caused by their transits are likely to be relatively shallow compared to the scatter of their light curves, and b) they are likely to have long orbital periods, so we can only detect at most a handful of transits in a 4 year window. These effects imply that Earth-like planets are more likely to have signal-to-noise ratios and multiple event statistics near or below the detection limit of the \Kepler\ pipeline. The probability of the \Kepler\ pipeline missing these planets is high -- it recovers only 26\% of injected transits\footnote{Calculated as described in Section 4 of \citet{dr25-kp-efficiency}. Also see Figure 7 in \citet{christiansen3}.} with expected MES between 7.1 and 8, despite these signals being strong enough (in principle) to rule out being fluctuations of pure random noise. There are many reasons that real transiting planets above the MES threshold of 7.1 can be rejected: residual non-Gaussian noise (like systematics and cosmic ray hits) can artificially reduce the apparent MES, light curve detrending can make transits appear shallower, and quantization in the grid search can cause slight mismatches in the detected period, epoch, and transit duration \citep{dr25-kp-efficiency}.

We can recover some of these lost planets by reducing the signal-to-noise threshold of the detection pipeline (e.g. \Kepler\ pipeline or BLS algorithm) to allow less-significant signals to become TCEs. As we showed in Section \ref{newcandidates}, this increases the recovery rate of planets at signal-to-noise {\em above} more conservative thresholds, but at the cost of detecting a large number of false positive TCEs that cannot practically be examined by eye. Therefore, a thorough study of the low signal-to-noise regime of \Kepler\ data would require an automatic vetting system that can accurately separate plausible planet candidates from false positives.

%as the MES threshold decreases, the probability of a detection caused purely by random noise increases
%Therefore, we must contend with a large number of random noise detections in addition to the types of false positives already discussed in this paper.
%Our ability to search for planets at low signal-to-noise is limited by our ability to accurately separate plausible signals from false positives.
% With a good enough automatic vetting system, we could find undiscovered treasures that have been hiding in plain sight in the \Kepler\ data.

Our neural network model for automatic vetting can quite accurately distinguish real planet signals from false positives caused by astrophysical and instrumental phenomena. 98.8\% of the time it ranks planet candidates higher than false positives in our test set. When testing our model on simulated data, we found that while our model recovers simulated true planets at a good rate, our current implementation does not yet match the performance of the more mature Robovetter system at identifying certain classes of simulated false positives, such as weak secondary eclipses and spurious detections. Nonetheless, when we applied our model to new \Kepler\ planet candidates, we found that it does a good job of ranking these TCEs by the probability of being a planet with very few obvious errors. We discovered several real new planets at the top of our model's rankings, which validates both the effectiveness of our model and our motivation for exploring the lower signal-to-noise region of \Kepler\ data. The immediate impact of this work is the discovery of several new \Kepler\ planets and planet candidates, including the exciting discovery of an eighth planet in the \Kepler-90 system.

At a higher level, a technique like ours could be used in the future to make more accurate estimates of planetary occurrence rates. In particular, the occurrence rate of Earth-like planets, colloquially called ``$\eta$-Earth'', is one of the most important and exciting open questions in exoplanet research -- it is directly proportional to the estimated fraction of planets that might harbor life as we know it on Earth \citep{drake1965}. A good measurement of $\eta$-Earth is crucial for designing the next generation of large space telescopes like LUVOIR and HABEX, which hope to directly image temperate, terrestrial worlds. To this end, we have identified the following ways that we can improve our model in the future to improve its accuracy and reduce known types of false positives:

\begin{enumerate}
    \item {\ron \textbf{Improved training set.}} Our current training set contains only about $15,000$ labeled examples, all of which are TCEs detected by the \Kepler\ pipeline with MES greater than 7.1. Only 8\% of planets and planet candidates in our training set have MES less than 10, and as a result we see a significant drop off in {\ron our model's performance on real planets and planet candidates} at MES $\lesssim 10$ (see Figure \ref{inj1}). Incorporating simulated data or unlabeled data into our training set would significantly increase its size, and would likely improve the performance of our model on low-MES planet candidates, eclipsing binaries and spurious false positives (see Section \ref{simdata}).
    \item {\ron \textbf{Improved input representations.}} Our current method for flattening light curves occasionally creates bogus transits in stars with high-frequency stellar variability (see Section \ref{simdata}). Improving our flattening routine could cut down on the number of signals due to stellar variability that are classified as likely planets. 
    \item {\ron \textbf{Centroid information.}} Incorporating some form of centroid information into our input representation may improve our model's ability to classify transits that occur on a background star instead of the target star. This could either be a phase-folded time series of RA/Dec (and some information about nearby stars and the expected centroid motion) or an in-transit difference image. 
    \item {\ron \textbf{Segmented local views.}} Splitting the local views into different segments would enable our model to examine the consistency of transits between different slices of the dataset. For example,  ``even-numbered transits'' and ``odd-numbered transits'', ``first half of the dataset'' and ``second half of the dataset'', or ``CCD module 1'', ``CCD module 2'', ``CCD module 3'' and ``CCD module 4''.
    \item {\ron \textbf{Secondary and tertiary local views.}} This would give our model a ``close up'' view of any putative secondary eclipses.
    \item {\ron \textbf{Auxiliary scalar features.}} We experimented with TCE features, such as period, duration, depth and impact, but we did not see any improvement on our validation set. We would like to try using stellar host features in the future, such as radius and density (to compare against the stellar density inferred from the transits).
    \item {\ron \textbf{Robust (outlier-resistant) means.}} The uncertainty on a sample median is $\sqrt{\pi/2} \approx 1.253$  times larger than the uncertainty on a sample mean, so in principle, by using robust means instead of medians in our input representation, we could boost the S/N the model sees by up to 25\%. 
\end{enumerate}

\section{Conclusion}

We presented a method for automatically classifying \Kepler\ transiting planet candidates using deep learning. Our neural network model is able to accurately distinguish the subtle differences between transiting exoplanets and false positives like eclipsing binaries, instrumental artifacts, and stellar variability. On our test set, our model ranks true planet candidates above false positives 98.8\% of the time. On simulated data, while it recovers simulated planets at a good rate, our model is not yet as effective as the more mature Robovetter system at rejecting certain types of simulated false positives, such as weak secondary eclipses. We are planning on improving our model in the future to address the failure modes we observed in simulated data, with the ultimate goal of using our model to calculate the occurrence rate of Earth-like planets in the galaxy.

We tested the ability of our model to identify new planet candidates by conducting a search of known \Kepler\ multi-planet systems. We conducted our search at a lower signal-to-noise threshold than traditionally used, allowing a high rate of spurious false positives in order to discover new candidates that were not recovered by previous searches. Our detection rate of almost one event per star would have produced too many candidates to examine manually in a full search of the $\sim 200,000$ \Kepler\ target stars. We used our model to rank the new candidates by likelihood of being a real planet and discovered several plausible planet candidates at signal-to-noise {\em above} the traditional threshold of 7.1. We statistically validated two of these new candidates: \Kepler-80~g, which is part of an interesting five-planet resonant chain, and \Kepler-90~i, which is now the eighth confirmed planet in the \Kepler-90 system. This paper brings \Kepler-90 into a tie with our Sun as the star known to host the most planets.

\acknowledgments
We thank Juliette Becker, Jessie Christiansen, Steve Bryson, Susan Thompson, Geert Barentsen, George Dahl Samy Bengio, and Martti Kristiansen for helpful conversations. We thank Susan Thompson for sharing her manuscript in advance of its submission, and we thank Jeff Coughlin for his expertise on column anomaly false positives in \Kepler\ data. {\ron We thank the anonymous referee for their thoughtful report and helpful suggestions.} A.V. acknowledges partial support from the NSF Graduate Research Fellowship, Grant No. DGE 1144152 and from NASA's TESS mission under a subaward from the Massachusetts Institute of Technology to the Smithsonian Astrophysical Observatory, Sponsor Contract Number 5710003554.  This research has made use of NASA's Astrophysics Data System and the NASA Exoplanet Archive operated by the California Institute of Technology, under contract with NASA under the Exoplanet Exploration Program. Some of the data presented in this paper were obtained from the Mikulski Archive for Space Telescopes (MAST). STScI is operated by the Association of Universities for Research in Astronomy, Inc., under NASA contract NAS5--26555. Support for MAST for non--HST data is provided by the NASA Office of Space Science via grant NNX13AC07G and by other grants and contracts. This paper includes data collected by the \Kepler\ mission. Funding for the \Kepler\ mission is provided by the NASA Science Mission directorate. This work was performed in part under contract with the California Institute of Technology (Caltech)/Jet Propulsion Laboratory (JPL) funded by NASA through the Sagan Fellowship Program executed by the NASA Exoplanet Science Institute.

Facilities: \facility{\Kepler}

%\bibliographystyle{apj}
%\bibliography{refs}

\clearpage
%\LongTables
\begin{deluxetable*}{cccccccc}
\tablewidth{0pt}
\tabletypesize{\scriptsize}
\tablecaption{Ranked List of New Kepler TCEs \label{classtable} }
\tablehead{\colhead{\Kepler\ ID} &
\colhead{Period} &
\colhead{$T_0$} &
\colhead{Duration} &
\colhead{Impact} &
\colhead{$R_p/R^\star$} &
\colhead{Predicted Planet} &
\colhead{Notes}
\\
\colhead{} &
\colhead{{\em days}} &
\colhead{$BJD_{\rm TDB}^\dagger$ - 2454833} &
\colhead{{\em hours}} &
\colhead{} &
\colhead{} &
\colhead{Probability} &
\colhead{}}
\startdata

11442793 & 14.44912434 & 2.200 & 2.704 & 0.777 & 0.0085 & 0.942 & {\em a} \\
8480285 & 10.91423130 & 1.507 & 4.809 & 0.746 & 0.0056 & 0.941 & \\
11568987 & 27.13809776 & 17.397 & 3.007 & 0.550 & 0.0091 & 0.920 & \\
4852528 & 14.64553642 & 5.453 & 1.636 & 0.592 & 0.0145 & 0.916 & {\em b}\\
5972334 & 6.02215719 & 4.013 & 1.859 & 0.877 & 0.0145 & 0.896 & \\
10337517 & 11.14567280 & 8.359 & 3.193 & 0.612 & 0.0077 & 0.860 & \\
11030475 & 4.74503899 & 3.396 & 2.470 & 0.606 & 0.0122 & 0.858 & \\
4548011 & 13.95128345 & 4.758 & 4.092 & 0.661 & 0.0051 & 0.852 & \\
3832474 & 29.45780563 & 0.826 & 11.117 & 0.661 & 0.0153 & 0.847 & \\
11669239 & 7.55533743 & 6.545 & 2.214 & 0.527 & 0.0081 & 0.788 & \\
10130039 & 96.10472107 & 11.038 & 5.548 & 0.484 & 0.0075 & 0.764 & \\
5351250 & 261.91778564 & 247.108 & 7.775 & 0.009 & 0.0202 & 0.752 & \\
5301750 & 5.18081331 & 1.166 & 1.423 & 0.901 & 0.0088 & 0.751 & \\
8282651 & 22.22732544 & 6.448 & 2.596 & 0.515 & 0.0138 & 0.731 & \\
6786037 & 21.14108849 & 20.541 & 5.479 & 0.964 & 0.0155 & 0.712 & \\
6690082 & 9.82555485 & 8.214 & 2.032 & 0.632 & 0.0084 & 0.711 & \\
9896018 & 5.55300617 & 0.074 & 3.037 & 0.667 & 0.0087 & 0.707 & \\
11450414 & 7.87495708 & 2.770 & 1.533 & 0.709 & 0.0085 & 0.655 & \\
6021275 & 5.92093372 & 5.056 & 1.791 & 0.505 & 0.0043 & 0.618 & \\
4851530 & 206.21182251 & 62.663 & 5.313 & 0.375 & 0.0467 & 0.613 & \\
8804283 & 7.70983267 & 2.427 & 3.593 & 0.645 & 0.0068 & 0.604 & \\
3323887 & 39.06724930 & 15.271 & 5.383 & 0.923 & 0.0583 & 0.595 & \\
6508221 & 4.11963272 & 3.461 & 1.542 & 0.441 & 0.0063 & 0.595 & \\
9006186 & 51.94103241 & 34.969 & 3.557 & 0.854 & 0.0112 & 0.585 & \\
12061969 & 14.10193443 & 13.709 & 2.306 & 0.047 & 0.0192 & 0.579 & \\
11074178 & 4.88644457 & 1.973 & 2.691 & 0.752 & 0.0095 & 0.560 & \\
8397947 & 5.14999104 & 0.281 & 4.094 & 0.629 & 0.0107 & 0.516 & \\
11968463 & 3.37368155 & 1.139 & 2.512 & 0.850 & 0.0090 & 0.509 & \\
10600261 & 7.49254370 & 6.303 & 1.909 & 0.771 & 0.0089 & 0.507 & \\
3323887 & 38.83024216 & 20.515 & 9.247 & 1.022 & 0.1017 & 0.507 & \\

\enddata
\tablenotetext{$\dagger$}{ $T_0$ is measured in Barycentric Julian Day (BJD), minus a constant offset of 2,454,833.0 days. The offset corresponds to 12:00 on Jan 1, 2009 UTC. The time given is the center of the first transit occurring after the reference date.}
%, modulo the detected period. % -- it's the same if we just say it's the first transit after the reference date as saying 'mod the period' right? 
\tablenotetext{$a$}{\Kepler-90 i.}
\tablenotetext{$b$}{\Kepler-80 g.}

\end{deluxetable*}

\clearpage

\end{document}